\documentclass[a4paper]{article}
\usepackage{jheppub}

\usepackage[utf8x]{inputenc}
\usepackage[T1]{fontenc}

\usepackage{amsmath}
\usepackage{amssymb}
\usepackage{amsfonts}
\usepackage{graphicx}
\usepackage[colorinlistoftodos]{todonotes}
\usepackage{soul}

\usepackage{booktabs}
\usepackage{siunitx}

\usepackage{booktabs}
\usepackage{array}
\newcolumntype{L}[1]{>{\raggedright\arraybackslash}p{#1}}

\usepackage{tikz}
\usetikzlibrary{fit,positioning}

\newcommand{\be}{\begin{equation}}
\newcommand{\ee}{\end{equation}}

\newcommand{\bea}{\begin{eqnarray}}
\newcommand{\eea}{\end{eqnarray}}
\newcommand{\Rep}{{\rm Re}\,}
\newcommand{\Imp}{{\rm Im}\,}

\newcommand{\vol}{\mathcal{V}}
\newcommand{\Mp}{M_{\rm P}}
\newcommand{\MC}[1]{\mathcal{#1}}

\begin{document}

\begin{flushleft}
DESY 18-177
\end{flushleft}

\title{Inflation as an Information Bottleneck: \\ A strategy for identifying universality classes and making robust predictions}

\author[a]{Mafalda Dias,}
\emailAdd{mafalda.dias@desy.de}
\author[a]{Jonathan Frazer,}
\emailAdd{jonathangfrazer@gmail.com}
\author[a]{Alexander Westphal}
\emailAdd{alexander.westphal@desy.de}

\affiliation[a]{Deutsches Elektronen-Synchrotron DESY, Notkestra{\ss}e 85, 22607 Hamburg, Germany}

\abstract{
In this work we propose a statistical approach to handling sources of theoretical uncertainty in string theory models of inflation. By viewing a model of inflation as a probabilistic graph, we show that there is an inevitable information bottleneck between the ultraviolet input of the theory and observables, as a simple consequence of the data processing theorem. This information bottleneck can result in strong hierarchies in the sensitivity of observables to the parameters of the underlying model and hence universal predictions with respect to at least some microphysical considerations. We also find other intriguing behaviour, such as sharp transitions in the predictions when certain hyperparameters cross a critical value. We develop a robust numerical approach to studying these behaviours by adapting methods often seen in the context of machine learning. We first test our approach by applying it to well known examples of universality, sharp transitions, and concentration phenomena in random matrix theory. We then apply the method to inflation with axion monodromy. We find universality with respect to a number of model parameters and that consistency with observational constraints implies that with very high probability certain perturbative corrections are non-negligible. 

}

\maketitle


\section{Motivation}


Cosmological inflation~\cite{Guth:1980zm,Linde:1981mu,Albrecht:1982wi} is currently our most promising candidate description of the very early Universe. But while precision cosmological observations are in striking compatibility with the simplest inflation models consisting of a single scalar field with a slow-roll flat potential, 
the relation between such a period and high energy physics remains unclear. 
A key element of this difficulty is the fact that slow-roll inflation is sensitive to ultraviolet (UV) physics.  Higher-dimension operators arising from radiative corrections, suppressed couplings to heavy fields and quantum gravity itself generically contribute ${\cal O}(1)$ corrections to the inflaton mass, spoiling the flatness of the potential. Inflation therefore requires a description within a candidate UV complete theory of quantum gravity such as string theory. Within string theory, the problem of UV sensitivity expresses itself in a particularly explicit form: string theory compactifictions to four dimensions have at minimum a spectrum of light moduli fields as well as KK modes. The couplings of a candidate inflaton to these scalar fields will generically generate at least a finite number of dangerous dimension-6 operators which have a drastic impact on the scalar potential, and of which we have little explicit theoretical control.\footnote{In certain cases however~\cite{Baumann:2007ah,Baumann:2010sx,Gandhi:2011id,Agarwal:2011wm}, these corrections are amenable to structural analysis using holographic methods, which in turn allows to treat their Wilsonian EFT coefficients as a finite set of random variables.} 

Given this situation, clear statements about the predictions of inflation models in string theory seem to be elusive. An inflation model in string theory will generally have multiple, and possibly a very large number of, free parameters. Some of these will represent the basic building blocks of the model, while others will come from various classes of subdominant corrections to the potential. While observable cosmological parameters can be extremely sensitive to such corrections, the range of possible corrections and consequent outcomes for a given inflationary scenario can seem somewhat overwhelming. Sometimes it is unclear whether or not it is possible even in principle for a model to be ruled out by observation.

In this work we want to address this difficulty by presenting a statistical approach for handling sources of theoretical uncertainty in string theory models of inflation. Our strategy is based on a result from information theory known as the  data processing theorem, with the key observation being that an \emph{information bottleneck} occurs whenever we coarse-grain a large amount of a priori UV information into a few observables, such as the amplitude of the power spectrum, its tilt, and the fractional power of inflationary gravitational waves. As a consequence of this information loss we do not expect observables to be equally sensitive to all UV parameters, and in particular we expect them to display \emph{universality} with respect to variations in some aspects of the UV input.
This behaviour is extreme in high-dimensional probability (a particularly well known subfield being random matrix theory), where models consisting of a large number of random variables very often demonstrate emergent simplicity as the model approaches some form of `large $n$' limit. This emergent simplicity, as well as the occurrence of sharp predictions (such as in the law of large numbers and concentration phenomenon more generally), are intimately related to universality and the data processing theorem. In this work we will find analogous phenomena in a more general setup, without requiring the model to necessarily be `large $n$'; the sacrifice we make for this significant practical gain is that our approach is purely numerical.

The application of these phenomena  to high energy physics is not new. There is a rather striking analogy between the very rich set of UV input possibilities for inflation and the small number of inflationary observables --- which are measured at long wavelengths of the curvature perturbations, so in the infrared (IR) --- and the structure of Wilsonian effective field theory (EFT) under renormalisation group flow. The most generic effective action we can write down for a given set of symmetries in the UV will have an infinite number of coupling constants, but under the renormalisation group this effective action flows to a theory with a finite set of marginal and relevant couplings in the deep IR. Analogies between inflation and renormalisation group flow have been noticed before, see \emph{e.g.}~\cite{ McFadden:2010na, Dias:2011in, Easther:2011wh, Kiritsis:2013gia, Binetruy:2014zya}. However the most striking parallel to us appears to be that both inflation and the renormalisation group flow for Wilsonian EFTs are realisations of the data processing theorem.

These phenomena suggest a more optimistic perspective on the complexity of inflation models coming from string theory. It may be that this rich UV input, rather than a hindrance to making predictions, can instead be seen as an opportunity enabling precise statements about the robustness of a given model to various sources of theoretical uncertainty\footnote{The beautiful lecture notes by Denef~\cite{Denef:2011ee} on complex structure make much the same argument.}. 
Ideally we would identify families of models of particular interest and establish clear probabilistic statements about their predictions, by identifying universality classes. 
With this aim we need to identify parameterisations of classes of contributions to the effective action (\emph{e.g.} perturbative corrections, instantons, or flattening mechanisms \dots), and then compute observables over the full parameter space. By embracing the full complexity of the model, we will inevitably see significant information loss between observables and the UV input. 
Understanding precisely how this information loss takes place can then be taken advantage of in understanding the conditions under which predictions are universal, or not, and hence direct further model refinement efforts by identifying which microphysical considerations have the greatest impact on observables. While we will focus on the case of inflation, the strategy we will present in this paper has a much broader scope of applicability, being potentially useful for any complex probabilistic model.

To illustrate our proposed method, in this paper 
we will focus on the model class defined by axion monodromy inflation in string theory~\cite{Silverstein:2008sg,McAllister:2008hb,Kaloper:2008fb,Dong:2010in,Kaloper:2011jz,Kaloper:2014zba,Palti:2014kza,Marchesano:2014mla,Hebecker:2014eua,McAllister:2014mpa,Ibanez:2014swa, Hebecker:2014kva,Hebecker:2015tzo,Landete:2016cix,Bielleman:2016grv,Bielleman:2016olv,Landete:2017amp}. 
Axion monodromy is a large field model with a weakly broken effective shift symmetry which controls an infinite series of corrections both perturbative and non-perturbative arising from the string theory embedding. Some of these corrections can potentially spoil inflation, while some others may generate distinctive features like flattening of the asymptotic monomial power of the scalar inflaton potential, $V\sim\phi^p$, or oscillatory patterns. As we will see, even such a specific model class can give rise to a broad range of outcomes for observables. 
Hence, the goal of our work is to identify the information bottleneck compressing the UV sensitivity of axion monodromy inflation into the IR observables and find universality classes --- identify the conditions under which we can make robust predictions. 

We will begin this endeavour by providing a detailed description of inflation as a probabilistic graph in \S\ref{sec:PGM}. In \S\ref{sec:ITandRMT} we introduce some quantities from information theory which we use as diagnostic tools and then apply them to studying well known phenomena in random matrix theory. This serves both as a means to test our numerical methods, as well as to provide a well understood example of the types of behaviour we would like to uncover in a model of inflation. In the remaining sections, we apply the same methods to axion monodromy inflation. We review the model in \S\ref{sec:model}, and then apply the analysis methods first to axion monodromy inflation without corrections in \S\ref{sec:results_tree}, and then to an extension of the model which includes parameterisations of `known unknowns' in \S\ref{sec:results_full}. Specifically, we include a contribution thought to account for all known lowest order perturbative corrections, as well as a broad class of non-perturbative corrections. For computational efficiency, our analysis of axion monodromy inflation also involves some machine learning to learn the function mapping the input parameters onto the observables $(P_\zeta,n_s,\alpha_s,r)$. This is described in \S\ref{sec:results_tree}. We discuss our results in \S\ref{sec:conclusions}.


From here on we set the Planck mass to unity. Also, for simplicity in the notation, we will not have distinct symbols for random variables (often denoted with upper case symbols) and particular instances of that random variable (often given the corresponding lower case symbol) as hopefully it will always be clear from the context which is being referred to.

\section{Describing inflation as a probabilistic graph}\label{sec:PGM}

The objective of this work is to build probabilistic models of inflation in order to better understand the map between model parameters and predictions, and make probabilistic statements about the outcomes for observables. As such, it is of central importance to go beyond scanning model parameters and instead view them as stochastic variables.
As proposed in Ref.~\cite{Price:2015qqb}, a suitable approach to doing so is to view a model of inflation as a probabilistic graph\footnote{Often also referred to as a Bayesian network, although in this work we will not be performing Bayesian inference. This would however be a very natural next step.} (PG) --- a probabilistic model which may be represented by a directed acyclic graph, where nodes correspond to model parameters and edges indicate either stochastic or deterministic dependencies.  
\begin{figure}
  \centering
     \includegraphics[width=0.45\textwidth]{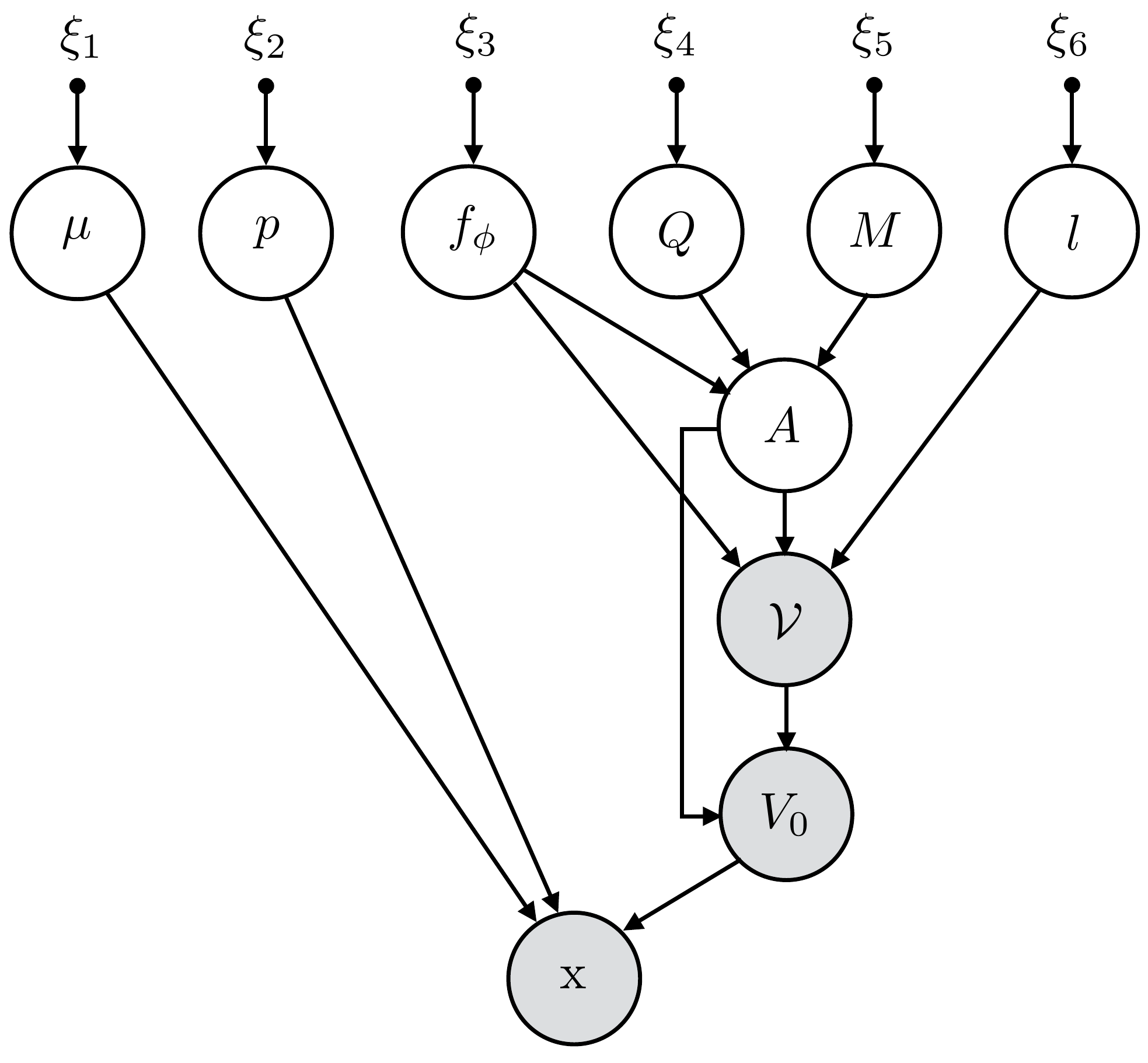}
  \caption{A simple graphical model of axion monodromy, corresponding to the tree level potential studied in \S\ref{sec:model}.}
  \label{fig:AM_graph_simple}
\end{figure}

An example of a PG, for a simple version of the inflation model studied in \S\ref{sec:model}, is shown in Fig.~\ref{fig:AM_graph_simple}. At the bottom, we have observables collectively denoted ${\bf x}$, which for the purpose of this work we take to be the amplitude of the power spectrum $P_{\zeta}$, the spectral index $n_{s}$, its running $\alpha_{s}$, and the tensor-to-scalar ratio $r$, such that ${\bf x} = \left\{P_{\zeta}, n_{s}, \alpha_{s}, r \right\}$
\footnote{In practice these are inferred quantities from cosmological surveys. Here we simply assume that the values these quantities take are known up to a given accuracy.}. The next layer up corresponds to a set of variables which represent model parameters in the usual sense, \emph{i.e.} parameters appearing in the action, in this case $\{V_0, \mu, p\}$. 
These parameters, or the distributions from which they are drawn, might be determined by other stochastic variables, which in turn can be drawn from a distribution determined by other stochastic variables. The arrows indicate the dependencies between the nodes and in principle there can be an arbitrary number of layers with any non-cyclic structure of dependencies (see for instance, figure 1 of Ref.~\cite{Leistedt:2018hmi} for another example, and figure 3 of Ref.~\cite{Feeney:2017sgx} for a more complex one). Here we use non-shaded nodes to represent stochastic dependencies and shaded grey nodes to indicate deterministic dependencies; for example, the observables represented by the ${\bf x}$-node are fully determined by the parameters showing up in the action.
Finally, at the very top layer of the graph we find the hyperparameters $\bf \xi$, deterministic variables which parameterise the distributions from which the model's stochastic variables $\bf z$ are drawn. For instance, if some $z_{i}$ were to be modelled as being drawn from the normal distribution $z_{i}\sim {\mathcal N}(\mu_{\mathcal N}, \sigma_{\mathcal N})$, then the mean $\mu_{\mathcal N}$ and standard deviation $\sigma_{\mathcal N}$ would be a natural choice of hyperparameters.

By clearly identifying dependencies through a network, this graphical language can be a useful tool to study models built within a high-energy physics embedding such as string theory. 
Deterministic dependencies such as those defining ${\bf x}$ are derived from the action and other top-down model building considerations. As an example, in the case of slow-roll single field inflation, we have the well-known expressions for $\bf x$ written in terms of slow-roll parameters $\epsilon, \eta$ and $\xi$ evaluated at the time the pivot scale leaves the horizon, $\phi_*$:
\be
P_{\zeta} = \left. \frac{H^2}{8 \pi^2 \epsilon}\right|_{\phi_*}  = \left. \frac{V^3}{12 \pi^2 {V^\prime}^2}\right|_{\phi_*}
\label{eq:Pz}
\ee
\be
n_{s} = \left[1 - 6\epsilon + 2 \eta \right]_{\phi_*} = \left[1 - 3\frac{V^{\prime 2}}{V^2} + 2\frac{V^{\prime\prime}}{V}\right]_{\phi_*}
\label{eq:ns}
\ee
\be
\alpha_{s} = \left[16 \epsilon\eta-24\epsilon^2-2\xi \right]_{\phi_*}= \left[8 \frac{V^{\prime 2}V^{\prime\prime}}{V^3} - 6 \frac{V^{\prime 4}}{V^4} - 2\frac{V^{\prime\prime\prime}V^\prime}{V^2} \right]_{\phi_*}
\label{eq:as}
\ee
\be
r = \left.16 \epsilon \right|_{\phi_*}= \left. 8\frac{V^{\prime 2}}{V^2} \right|_{\phi_*} \ ,
\label{eq:r}
\ee
where $V$ and $V^\prime$ are the potential and its gradient. Stochastic dependencies provide a means to handle sources of uncertainty, be it through having only partial knowledge of some aspects of the underlying physics, or because the physics is fundamentally stochastic. For example, as will be discussed in \S\ref{sec:model}, in this particular model the prior on $A$ is parameterised by $Q$ and $M$. We have some valuable information about these parameters, such as certain conditions they must satisfy and their order of magnitude but their precise value is unknown. We therefor parameterise our ignorance by treating them as stochastic variables drawn from a parametric distribution with hyperparameters which we can vary. Ultimately, by following the network of dependencies, the hyperparameters parameterise all the distributions of the model.  

The joint PDF $p({\bf z}; \xi)$, just like the deterministic dependencies, should also be derived from top-down microphysical considerations. At first, this step may seem like a hopeless task. 
One may be concerned that there is no systematic way in which to derive $p({\bf z}; \xi)$, making the whole model and its predictions rather meaningless. One of the central claims of this work is that all hope should not be lost, and that indeed string theory models of inflation are well equipped to make strong statements about the form of $p({\bf z}; \xi)$. The basic idea is that the series of deterministic or stochastic dependencies connecting a given fundamental parameter to an observable, can be viewed as a form of data processing, between the UV physics and the observables. So while there may be considerable uncertainty in the underlying physics, the structure of the graph will cause information loss and hence a loss of sensitivity to variations of at least some of the most fundamental parameters. If the information loss is severe, then this results in universality  with respect to a given parameter --- predictions that are totally insensitive to certain aspects of the underlying physics. This point will be discussed in detail in \S\ref{sec:ITandRMT}. Thus, what may at first seem like only very limited input from string theory considerations can turn out to be sufficient to make robust statements about the prediction of the model. 

In practice, what we will see is a tendency for strong hierarchies between the parameters. Hence, there will be universality with respect to some, possibly many parameters, but there will be a subset of parameters to which predictions are sensitive. This situation is reminiscent of identifying relevant couplings under renormalisation group flow. Although in this case, we do not have the luxury of a scale parameter and will therefore need to discover the hierarchies using numerical methods.
To do so, we will make use of tools from information theory and machine learning. 
The strategy we will employ has three steps.

\subsection{Three step procedure for studying a probabilistic model of inflation}
%

\textbf{\emph{Definition of PG and identification of relevant scales --- }}  The first step is expressing the inflation model as a PG, where all parameters and hyperparameters are identified and their dependencies appropriately defined. As discussed, it is often not easy to compute the distributions for stochastic parameters from top-down considerations, but it is however often possible to identify the relevant scales that each parameter can take. We use these bounds to build what we have decided to call \emph{fiducial} priors. Fiducial priors are built using micro-physical considerations or, in the lack of any top-down information, modelled assuming maximum ignorance, for example with a flat or log-flat distribution. We will then test the sensitivity to these assumption by studying variations of the prior relative to this fiducial model (this is the third step).


\textbf{\emph{Learning the mapping from model parameters to observables --- }} Once the model parameters and their fiducial priors and hyperparameters are fully defined, we want to compute the distributions for observables for a range of values of the hyperparameters. For this we use, for example in the case of slow-roll single-field inflation, the mapping defined by equations~\eqref{eq:Pz}-\eqref{eq:r}. 
For more complicated models involving more fields or deviations from slow-roll, while this mapping remains deterministic, rarely can one derive analogous expression and generally one must resort to numerics. Indeed, even in situations where  expressions~\eqref{eq:Pz}-\eqref{eq:r} do hold, it is still often the case that there is some need for numerics. The reason for this is simply that these expressions require knowledge of $\phi_{*}$ ---  the point in field-space corresponding to the horizon-crossing of the relevant pivot-scale --- and even in rather simple models it can be challenging to find analytical expressions valid over the full parameter space. 
The numerical alternative is to solve a system of ordinary differential equations for the inflationary trajectory, and if need be, for the perturbations as well. This must be repeated for each random draw of the model parameters and hence computing distributions for observables is a computationally heavy task. Indeed, the numerical expense of computing these distributions for a large range of hyperparameters can quickly become inaccessible.

To bypass this problem, we appeal to basic machine learning methods. The idea is to build the distribution for observables using equations~\eqref{eq:Pz}-\eqref{eq:r} only once, for the fiducial prior. We then use regression methods to \emph{learn} the mapping from parameters appearing in the action to observables, ${\bf x}(V_0,\mu,p)$. Once this function is built numerically, it is trivial to study how the distributions for observables change with hyperparameters. Exactly which regression method is best suited to learning this map is a model-dependent question. In section \S\ref{sec:model} we describe how we did this for the specific model of Fig.~\ref{fig:AM_graph_simple}, as well as for a more complete (and complex) version of this model. 

\textbf{\emph{Testing prior sensitivity with information theory --- }} Once we know how the distributions for observables change with variations of the hyperparameters, we would like to have a systematic and quantitative means to measure information loss, and hence identify hierarchies in dependencies and the emergence of universality. As the name might suggest, information theory has a number of suitable diagnostic tools. In particular, we will use the relative entropy (\emph{a.k.a} the Kullback--Leibler divergence) and mutual information, as we will now review. We use these tools to look for three specific traits --- robustness to prior assumptions, the presence of sharp transitions in predictions and we are also interested in the predictive power of the model.





\section{Elements of information theory}\label{sec:ITandRMT}

In this section we briefly introduce the information theoretic concepts motivating our approach to studying probabilistic models of inflation and more specifically, our choice of diagnostic tools for identifying certain model behaviours. The methods of this paper seek to be applicable to both simple and complex models alike, and as this paper hopefully demonstrates, they are useful for revealing non-trivial structure even in models with a small number of parameters. However, for more complex models we can expect these tools to be especially valuable. 

And yet, in many regards the behaviour of a probabilistic model will typically simplify once the model is sufficiently complex. A number of phenomena we are interested in are most easily understood in some form of large $n$ limit. For this reason we guided our analysis by making systematic contact with the wealth of results that have been developed in the context of high-dimensional probability. Despite our axion monodromy model not being high-dimensional in any sense, we found this approach to be fruitful, since at a qualitative level we identified similar behaviour. Hence, in this section, in addition to introducing some quantities from information theory, we also introduce some ideas from high-dimensional probability. We then use simple examples with large random matrices to illustrate the utility of our information theoretic tools in identifying this behaviour.

As beautifully described by van Handel~\cite{van2014probability}, there is a list of general principles which are frequently applicable in the study of high-dimensional probability: concentration of measure, suprema (as a means of estimating expectation values), universality, and sharp transitions. In this work we are primarily concerned with understanding the conditions under which a model of inflation gives robust predictions, hence methods relating to universality are of central importance. Concentration of measure is clearly another important topic in this context since it presents the tantalising possibility that not only can predictions be robust when the model is sufficiently complex, but also that the predictions may be sharp. Since we are primarily concerned with numerical methods in this work, the usage of suprema for estimating expectation values is not really required. Sharp transitions are a fascinating aspect of large-$n$ behaviour and we will see that the information measures we use here lend an interesting new perspective to this topic.

A rather stunning demonstration of these basic principles of high dimensional probability arrises in the study of large random matrices. For the purpose of this work, we can view universality in random matrix theory (RMT) as a specific example of the sort of hierarchies we wish discover through numerical studies of probabilistic models of inflation. So after we introduce our information theory diagnostic tools, we will turn our discussion to a special class of random matrices known as Wigner matrices and use them both to demonstrate some of these ideas in action and also to explore the extent to which the information theoretic measures are effective as numerical probes of these concepts. We will then in \S\ref{sec:results_tree} and \S\ref{sec:results_full} apply the same principles to a model of axion monodromy (AM), first in its simplest tree level form, and second to a more complete formulation. Table \ref{table:roadmap} summarises some analogies between behaviour seen with random matrices and with axion monodromy, together with an example diagnostic tool. This table may also serve as a high level map to the rest of this paper.

\begin{table}[htbp]
\caption{Roadmap: Explore 3 general behaviours of probabilistic models in RMT and axion monodromy}
\begin{tabular}{@{}p{0.22\textwidth}*{4}{L{\dimexpr0.256\textwidth-2\tabcolsep\relax}}@{}}
\toprule
    & {RMT example} & {AM (tree level)} & {AM (with corrections)} \\ 
 \midrule
    {Universality}  & p($\lambda)$ & p($\bf x$) w.r.t. $\mu$ ; p($A$) & p($\bf x$) w.r.t. $C_n$ and $\chi_n$   \\
    {Predictivity} & p($\lambda_1)$ as $n \rightarrow \infty$  & p({\bf x}|$p$) & p($\bf x$) for small $V_0$   \\ 
    {Sharp transitions} & p$(\lambda_1)$ under shift of mean of $M_{n}$  & --- & p(${\bf x}\vert V_0)$ as $V_0$ changes  \\ 
\bottomrule
\end{tabular}
\label{table:roadmap}
\end{table}


\subsection{The diagnostic tools --- $f$-divergences and mutual information}\label{subsec:it}

Minimally adapting discussion in Ref.~\cite{ay2017information}, we can think of a model of inflation as a family of probability measures $\bf{p}$ on the space of parameters $\Omega$. We can see $\bf{p}$ as a mapping from the space of hyperparameters $M$ to probability density functions on $\Omega$,
\be
{\bf p}: M \rightarrow \mathcal{P}(\Omega) \ ,
\ee
such that we have a probability measure p$({\bf z};\xi)$ on the parameter space $\Omega$. 

Given such a model, we can then compute observables. This amounts to deriving a map from the parameter space $\Omega$ to the space of possible outcomes for observables $\Omega'$
\be
\kappa : \Omega \rightarrow \Omega'\, .
\ee
The map $\kappa$ induces a family of measures on $\Omega'$, which we denote $\bf{p}'$ and hence the statistical model $(M, \Omega, {\bf p})$ gets transformed into a model on observable space $(M, \Omega', {\bf p}')$.

If $\Omega'$ is smaller than $\Omega$ then in general we can expect information to be lost. Our goal is to study the characteristics of this information loss in order to assess universality and the robustness of the predictions of the model. There are many ways one might wish to do this but here we will primarily consider just two approaches. One way is to study the mapping $\kappa$ directly by computing how much information about a parameter $z_i$ is contained in an observable $x_j$; this is usually characterised by the mutual information. A different approach is to assess how moving around in the hyperparameter space is reflected in the induced model $(M, \Omega', {\bf p}')$, \textit{i.e.} find a measure of the distinguishability of points in M when viewing $(M, \Omega', {\bf p}')$; for this we can use the ratio of appropriately chosen $f$-divergences. The mutual information and $f$-divergences have different properties and provide complementary perspectives of the of the model. There are also situations where one quantity can be trivial to compute and the other extremely difficult. We now discuss each in turn in more detail, and in the following subsections we will apply them to random matrix theory examples which illustrate how they can be used to assess universality \S\ref{subsec:universality}, predictivity \S\ref{subsec:predictivity} and sharp transitions \S\ref{subsec:sharp}. 


\paragraph{Basic Properties of $f$-divergences} In this context, divergences can be thought of as measures of dissimilarity between probability distributions. However, it is important to note that in general they are neither symmetric, nor do they satisfy the triangle inequality and are thus not a distance measure in the usual sense. First introduced by Csisz\'ar~\cite{csiszar2008axiomatic}, the $f$-divergence between two distributions $p$ and $q$ is defined by
\be
D_{f}(p\Vert q) := \mathbb{E}_{q}\left [f\left ( \frac{p}{q}\right)\right ] \, , 
\ee
where $f$ is a strictly convex function $f: \mathbb{R}^{+}\rightarrow \mathbb{R}$, and $\mathbb{E}_{q}$ is the expectation value with respect to $q$. A particularly important example of an $f$-divergence is the choice $f(x) = x \ln x$, which is known as the Kullback -- Leibler divergence, otherwise known as the relative entropy. In this work we will focus on this case, which we denote $D_{\rm KL}(p\Vert q)$ \footnote{Other well known examples include total variation, $\chi^{2}$-distance, Hellinger-distance and Tsallis-distance. It should be noted however that ``not all $f$-divergences are born equal" and that a considerable body of work exists that seeks to find bounds on one divergence in terms of another, which may be more tractable in a specific scenario. We do not take advantage of this rich history here. Instead we will focus on building intuition about what can be learnt from $D_{\rm KL}(p\Vert q)$.}.  The key feature of a divergence, which makes it a ``distance-like" measure, is that it satisfies Gibbs' inequality:
\be
D_{f}(p\Vert q) \geq 0 \ ,
\ee
where $D_f(p\Vert q) = 0$ if and only if $p = q$. Another key property of $f$-divergences (and hence also of the relative entropy) is that they satisfy monotonicity under information loss (otherwise known as the Data Processing Theorem) \cite{csiszar2008axiomatic}:
\begin{quote}
If $\kappa$ is an arbitrary transition probability (deterministic or stochastic map) that transforms measure $p$ and $q$ into $p_{\kappa}$ and $q_{\kappa}$, then
\be\label{eq:DPT}
D_{f}(p\Vert q)\geq D_{f}(p_{\kappa}\Vert q_{\kappa})
\ee
where equality holds if and only if the transition is induced from a sufficient statistic with respect to $\{p,q\}$.
\end{quote}
The theorem refers to a Markov chain in describing $\kappa$ as a transition probability but here we are mostly concerned with the special case where $\kappa$ is deterministic. To understand the implications of this theorem consider $p$ to be the distribution of some parameter $z$ for a choice of hyperparameters $\xi_1$ --- $p(z;\xi_1)$ --- and $q$ to be $p(z;\xi_2)$, and $\kappa$ to be the map from $\Omega$ to $\Omega'$ described above. 
Expression~\eqref{eq:DPT} states that $D_{f}(p(z;\xi_1)\Vert p(z;\xi_2))\geq D_{f}(p(x;\xi_1)\Vert p(x;\xi_2))$, \textit{i.e.} there is information loss in the map from parameters to observables and this loss can be measured by the change in $f$-divergence between two points in $M$ under this map. $f$-divergences, and in particular the KL-divergence, are therefore useful quantities to assess universality and robustness.

Anticipating discussion of universality in subsequent sections, an important special case is when a subclass of model parameters ${\bf z} = (z_{1},\dots,z_{n})$ are statistically independent such that
\be
p({\bf z}) = \prod_{i}^{n} p(z_{i}) \, .
\ee
In this case the data processing theorem reads
\be\label{eq:DPT_indep}
\sum_{i}^{n} D_{\rm KL}\left (p({z_{i}})\Vert q({z_{i}})\right ) \geq D_{\rm KL}\left (p_{\kappa}(x)\Vert q_\kappa(x)\right) \, .
\ee
The proof is given in Appendix \ref{sec:DPT_proof}. Note that in the case where $z_{i}$ are not only independent but also identically distributed (iid), the left hand side of Eq.~\eqref{eq:DPT_indep} is simply $n D_{\rm KL}\left (p({z_{i}})\Vert q({z_{i}})\right )$. 
This motivates the following measure of information processing	
\be\label{eq:I_f_def}
\mathcal{I}_{f}(\xi_{1},\xi_{2})  := \frac{D_{f}(p(x;\xi_{1})\Vert p(x;\xi_{2}))}{D_{f}(p(z_{i};\xi_{1})\Vert p(z_{i};\xi_{2}))} \, .
\ee
This object takes values in the range $\mathcal{I}_{f} \in \left [0, n \right. ]$. For $\xi_{1}\neq\xi_{2}$, $\mathcal{I}_{f} = 0$ implies total universality and, at the other extreme, $\mathcal{I}_{f} = n$ implies precisely zero loss of information. For $n > 1$, the intermediate value $\mathcal{I}_{f} = 1$ seems to represent the boundary between two interesting classes of behaviour. If $\mathcal{I}_{f} < 1$ then clearly information is being lost and for models exhibiting universality in the large $n$ limit we can probe the approach to universality by studying the $n$-dependence of $\mathcal{I}_{f}$. $\mathcal{I}_{f} > 1$ implies that $x$ is a sensitive probe of $\xi$ and in many situations we can expect the sensitivity to improve with $n$. An extreme example is if $x$ is a sufficient statistic for $\xi$. For instance, if $x$ is the sample mean $x = \frac{1}{n}\sum_{i}^{n}z_{i}$ and the $z_{i}$ are drawn from a normal distribution with fixed variance, and the sole hyperparameter $\xi = \mu$ being the mean, then clearly the sensitivity to $\mu$ increases with $n$ and thus $\mathcal{I}_{f}$ increases from one monotonically with $n$. We will explore these ideas in more detail in \S\ref{subsec:sharp}.



\paragraph{Basic Properties of the Mutual Information}
The mutual information measures the average information a random variable $x$ conveys about another variable $z$ 
\be
I(x;z): = \int dx dz\, p(x,z)\log{\frac{p(x,z)}{p(x)p(z)}}\, ,
\ee
which is equivalent to the relative entropy between the joint probability $p(x,z)$ and the product of the marginal distributions $p(x)p(z)$
\be
I(x;z) \equiv D_{\rm KL}(p(x, z)\Vert p(x)p(z))\, .
\ee
Thus $I(x;z)\geq 0$ and the bound is saturated when $x$ and $z$ are statistically independent. Recently, the mutual information was successfully used to gain insight into the process of deep learning using the information bottleneck principle \cite{tishby2000information, tishby2015deep, shwartz2017opening}\footnote{The methods of Refs.~\cite{tishby2000information, tishby2015deep, shwartz2017opening} are not directly applicable to this work but they were clearly influential and hence we pay tribute with our choice of title for this paper.}. Given a graphical model such as the one given in Fig.~\ref{fig:AM_graph_simple}, we can compute the mutual information between nodes to diagnose hierarchies in dependencies.

\begin{figure}
  \centering
     \includegraphics[width=0.26\textwidth]{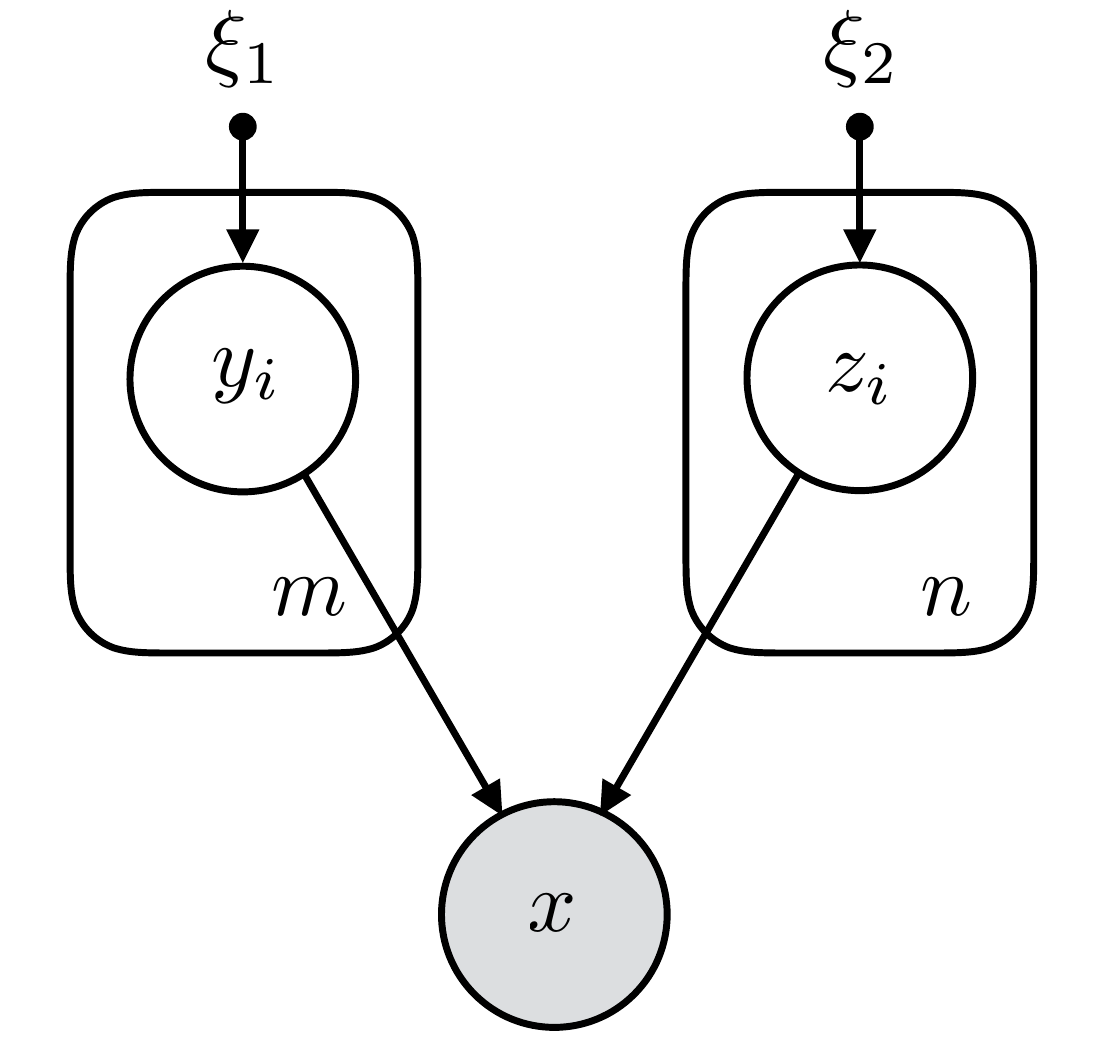}
  \caption{A model}
  \label{fig:two_classes_model}
\end{figure}

\paragraph{$I$ \emph{Vs} $\mathcal{I}_{f}$}
Both the mutual information and ratios of the relative entropy, or KL-divergence, can be used to study information loss in the mapping from parameter space to observable space but the precise questions they are suited to addressing differ. They also have different practical considerations. To illustrate this, consider a model of the form shown in Fig.~\ref{fig:two_classes_model}. The plates around variables $y_i$ and $z_i$ indicate that there are $m$ independent and identically distributed (iid) variables $y_i$, with the same hyperperameter $\xi_1$ and $n$ iid variables $z_{i}$ with hyperparamter $\xi_2$. 
One might have a number of questions about this hypothetical model. For instance:
\begin{enumerate}
\item How much information does $x$ contain about each of the $m + n$ latent variables?
\item How much information does $x$ contain about the variables ${\bf y}=(y_{1},\dots, y_{m})$ collectively \emph{vs.} the variables ${\bf z} =(z_{1},\dots,z_{n})$?
\end{enumerate}
The first question is naturally addressed by computing the mutual information between each latent variable and the observables --- $I(x;y_{i})$ and $I(x;z_{j})$. This can be done quite efficiently even for large $m+n$ and a number of publicly available packages exist (we tried the codes resulting from Refs.~\cite{DBLP:journals/corr/GaoSG14, szabo14information, szabo12separation}) which will compute these quantities directly from random samples (\emph{i.e.} bypassing the need to estimate the joint distributions $p(x, y_{i})$, etc.). The second question could also be studied with the mutual information, by computing $I(x; {\bf y})$ and $I(x; {\bf z})$. However, unless the model exhibits a symmetry which simplifies the problem, this calculation becomes computationally intensive for large $m + n$. The KL-divergence, on the other hand, is easy to compute and hence question 2 might be better addressed by computing $\mathcal{I}_{\rm KL}(\xi_1,\xi_1')$ and $\mathcal{I}_{\rm KL}(\xi_2,\xi_2')$, where $\xi_1 \rightarrow \xi_1'$ and $\xi_2 \rightarrow \xi_2'$ are coordinate changes in hyperparameter space. 
The key point to note here is that despite the fact that $x$ is the result of a map from an $(n+m)$-dimensional parameter space to a 1-dimensional observable space, this analysis takes advantage of the fact that there are only two classes of iid variables. So instead of studying the $(m+n)$-dimensional family of distributions on parameter space $p({\bf y},{\bf z};\xi_1, \xi_2)$, one can study information loss through sensitivity to changes in hyperparameter space. Hence instead of scaling with $m+n$ the study just scales with the number of hyperparameters, of which, in many situations, there are only a few. We will explore these possibilities further in the following sections.

\subsection{Universality} \label{subsec:universality}

Central to this paper is the task of identifying universality, which occurs whenever the distribution of a variable $x=f(z_1, \dots, z_{n})$ is robust against changes in the distributions of random variables $z_{1},\dots,z_{n}$. 
It is interesting to make a connection between the information theory language just discussed and approaches in high dimensional probability used to get analytic results. In this regard, a common choice for a measure of the robustness of the distribution on $x$ is the expectation value. In this case, universality can be informally defined as done in Ref.~\cite{van2014probability}:
\begin{quote}
If $z_{1},\dots,z_{n}$ are independent (or weakly dependent) random variables, then the expectation $\mathbb{E}[f(z_{1},\dots,z_{n})]$ is "insensitive" to the distribution of $z_{1},\dots,z_{n}$ when f is "sufficiently smooth".
\end{quote}
This statement can be made precise in a number of ways, as discussed in appendix~\ref{app:HDPRMT}.
At a first glance it seems like the requirement of Eq.~\eqref{eq:I_f_def}, \textit{i.e.} $\mathcal{I}_{\rm KL} \rightarrow 0$, is much more stringent than requiring constant expectation values. This is of course true, however, as will be shown in  appendix~\ref{app:HDPRMT}, studying the behaviour of full distributions can be recast as computing an expectation value of an appropriately chosen quantity. Thus this statement, and in particular Eq.~\eqref{eq:universality}, can in some cases be sufficient to prove complete universality analytically.
To our knowledge, results such as Eq.~\eqref{eq:universality} have never been applied to inflation models, however numerical evidence of universality in multifield inflation has been reported for handful of models~\cite{Easther:2013rva, Price:2015qqb, Dias:2017gva}.

An especially striking demonstration of universality arrises in the study of large random matrices, in particular in the study of the global properties of the eigenvalue spectrum of Wigner matrices. We present the Wigner ensemble using the definition in Ref.~\cite{tao2012topics}:
\begin{quote}
A Wigner Hermitian matrix ensemble is a random matrix ensemble $A_{n} = (A_{ij})_{1\leq i,j \leq n}$ of Hermitian matrices ([...]), in which the upper-triangular entries $A_{ij}$, $i > j$ are iid complex random variables with mean zero and unit variance, and the diagonal entries $A_{ii}$ are iid real variables, independent of the upper-triangular entries, with bounded mean and variance.
\end{quote}
In this work we construct numerically an ensemble of Wigner matrices by first constructing a non-symmetric $n \times n$ random matrix $M_{n}$ with entries $\Rep (M_{n})_{ij}\sim p_{M}$, $\Imp (M_{n})_{ij}\sim p_{M}$, where $p_{M}$ is a distribution with support over $\mathbb{R}$ and appropriately rescaled such that $\mathbb{E}[(M_{n})_{ij}] = 0$ and $\mathrm{Var}[(M_{n})_{ij}] = 1$. We then build a Hermitian matrix $A_{n}$ which belongs to the Wigner ensemble by defining
\be
A_{n} = \frac{1}{2}\left( M_{n} + (M_{n})^{\dagger}\right )\,.
\ee
With this definition, all entries have variance $\mathrm{Var}[(A_{n})_{ij}] = 1$. Note that the definition of a Wigner matrix above only restricts the first two moments of the entries of $A_{n}$ and thus we have considerable freedom in our choice of $p_{M}$.

\begin{figure}[t]
  \centering
     \includegraphics[width=0.95\textwidth]{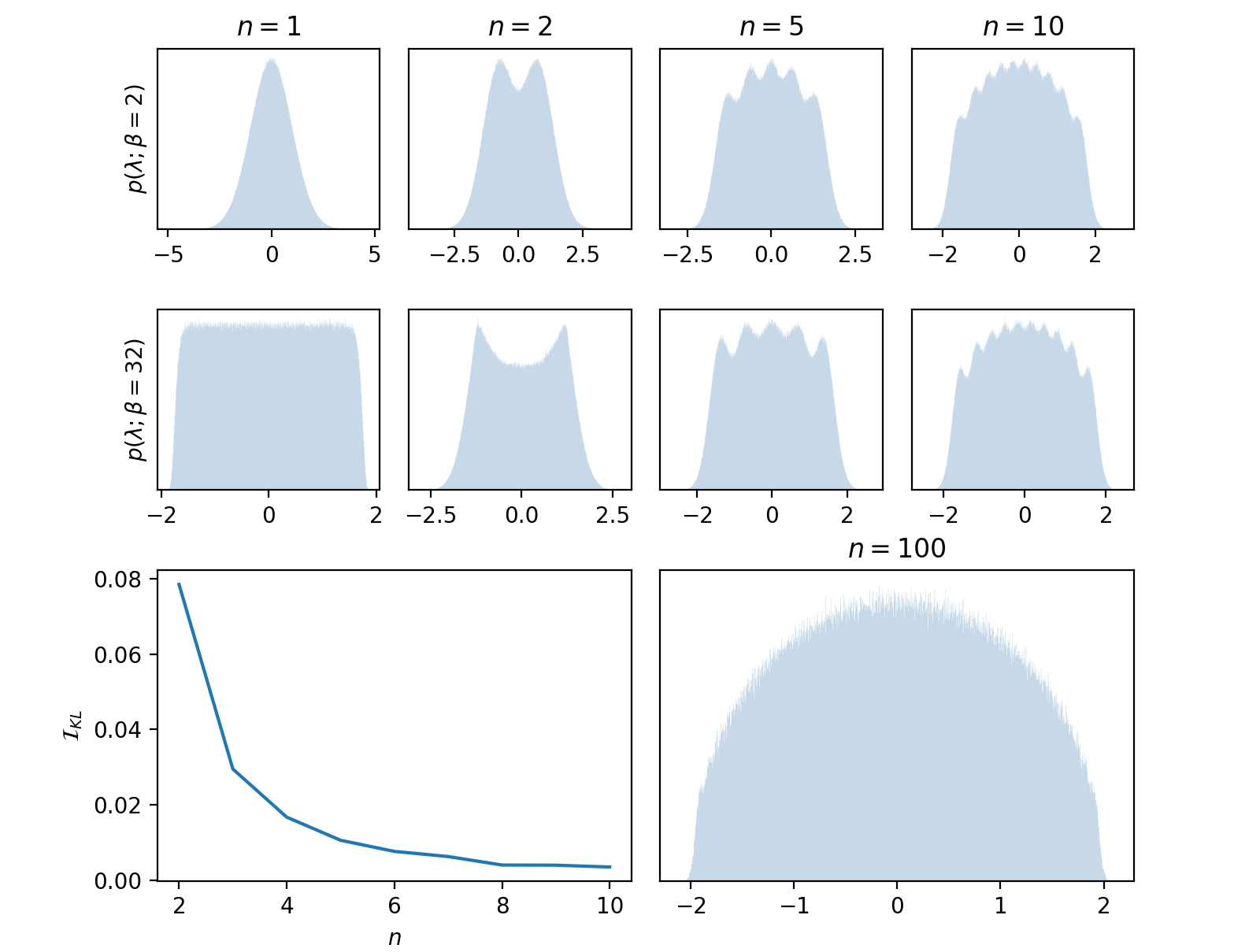}
\caption{The top two rows show histograms of the scaled eigenvalues $\lambda/\sqrt{n}$ of $n \times n$ Wigner matrices, 
with the real and imaginary components of $(M_{n})_{ij}$ drawn from the generalised normal distribution with hyperparameter $\beta =2$ in the top row and $\beta =32$ in the second row. We have rescaled the distribution to ensure that $\mathrm{Var}[M_{ij}] = 1$. The $n = 1$ case shows the shape of these two priors. As $n$ increases, the distributions of the eigenvalues become increasingly similar. A measure of this similarity is given by $\mathcal{I}_{\rm KL} = D_{\rm KL}(p(\lambda; \beta = 2)\Vert p(\lambda; \beta = 32))/D_{\rm KL}(p(M_{ij}; \beta = 2)\Vert p(M_{ij}; \beta = 32))$, shown in the bottom left plot. As $n$ increases $\mathcal{I}_{\rm KL} $ rapidly approaches zero. The bottom right plot shows a histogram of the eigenvalues of an ensemble of $100 \times 100$ Wigner matrices. This distribution closely resembles the Wigner semicircle law, proven to be universal in the large $n$ limit.}

  \label{fig:prior_sens_uni}
\end{figure}

To see the emergence of universality in the global statistics of this ensemble, consider Fig.~\ref{fig:prior_sens_uni}. The histograms are constructed by generating a large ensemble of Wigner matrices and then binning the resulting eigenvalues rescaled by a factor or $\sqrt{n}$ \footnote{The edges of the distribution scale as $\sqrt{n}$ so by studying the spectrum of $A_{n}/\sqrt{n}$ one makes the width of the distribution independent of $n$.}. Moving from left to right, the top two rows show the resulting eigenvalue spectrums from $1 \times 1$, $2 \times 2$, $5 \times 5$, and $10 \times 10$, matrices. The first and second rows differ in the choice of prior for $(M_{n})_{ij}$, the shape of which can be seen directly from the $1 \times 1$ case. Both distributions can be thought of as different hyperparameter choices of the generalised normal distribution $(M_{n})_{ij} \sim p(\cdot;\beta)$ with density function
\be
p(x; \beta) \sim \frac{\beta}{2\alpha \Gamma(1/\beta)}e^{-\left( \frac{\vert x\vert}{\alpha}\right )^{\beta}} 
\ee 
where $\Gamma$ is the usual Gamma function and we set $\alpha  = \sqrt{\Gamma(1/\beta)/\Gamma(3/\beta)}$ to ensure unit variance. When $\beta = 2$ the distribution is simply the normal distribution, which is the case for the top row of Fig.~\ref{fig:prior_sens_uni}. As $n$ increases, the histograms in the first and second row appear to become increasingly similar. The plot on the bottom left of Fig.~\ref{fig:prior_sens_uni} confirms this by computing $\mathcal{I}_{\rm KL} = D_{\rm KL}(p(\lambda; \beta = 2)\Vert p(\lambda; \beta = 32))/D_{\rm KL}(p(M_{ij}; \beta = 2)\Vert p(M_{ij}; \beta = 32))$, which quickly approaches 0 with increasing $n$. 
Indeed, as $n$ becomes large, regardless of the choice of prior, the distribution of eigenvalues of $A_{n}/\sqrt{n}$ converges to the famous Wigner semicircle law:
\be
p(\lambda) = \frac{1}{2\pi}\sqrt{4-\lambda^2} \, \delta_{|\lambda|\leq 2} \, ,
\label{eq:semi-circle}
\ee
where $\delta_{|\lambda|\leq 2}$ indicates $p(\lambda)=0$ for $|\lambda|>2$.
This is illustrated by the bottom right hand plot, which shows the spectrum for $100 \times 100$ matrices. This striking example of universality can be proven using standard tools in high-dimensional probability, as described in appendix \ref{app:HDPRMT}.

This rather brief example illustrates precisely the sort of behaviour we would like to discover in a given inflation model. From a model building perspective, if we imagine that the eigenvalue spectrum corresponds to an observable, then in order to make robust predictions, we only need very limited (but specific) knowledge about the statistics of the entries in order to make robust statements about the predictions. Specifically, in this case, we care about the mean and variance of the entries, but beyond that the ``predictions'' become robust to our choice of prior once $n$ is sufficiently large.

\subsection{Predictivity of the model, or concentration of measure} \label{subsec:predictivity}

Beside diagnosing if the predictions of a model are robust against considerations about the underlying construction, one can also study if or when these predictions become sharp. This phenomenon is referred to as concentration of measure in the context of high dimensional probability. Ref.~\cite{van2014probability} informally defines it as
\begin{quote}
If $z_{1},\dots,z_{n}$ are independent (or weakly dependent) random variables, then the random variable $f(z_{1},\dots,z_{n})$ is close to its mean $\mathbb{E}[f(z_{1},\dots,z_{n})]$ provided the function $f(z_{1},\dots,z_{n})$ is not too sensitive to any of the variables $z_{i}$.
\end{quote}
So provided certain criteria are met (which determines the precise meaning of "not too sensitive") then sufficiently complex models can be highly predictive. 

In the context of inflation, this idea has not been studied in detail, although some first steps have been taken. There are at least two ways in which large $n$ phenomena can arise in inflation. One can consider inflation with a very large number of fields (sometimes referred to as manyfield inflation), or one can consider a small number of fields but a large number of terms appearing in the effective potential. In the case of manyfield inflation, numerical studies of a specific model of $n$-flation in Refs.~\cite{Easther:2013rva,Price:2014ufa,Price:2015qqb} found that models consisting of $\mathcal{O}(100)$ fields had similar predictive power to single field models \footnote{More precisely, it was argued in Ref.~\cite{Price:2015qqb} that the dominant source of uncertainty was the choice of pivot scale as determined by reheating. A problem that impacts single field and multifield models equally.}. This is despite the fact that for a modest number of fields $1< n \lesssim 10$ the model is less predictive. More recently, similar studies have been carried out for considerably more complex manyfield models, showing that at least for some observables (but not all) models are again highly predictive \cite{Dias:2016slx, Dias:2017gva, Bjorkmo:2017nzd}. In the case of single field models with a large number of terms in the potential, cencentration of measure with regard to predictions has to our knowledge not been directly studied. However this was possibly a motivation for the work that was done in Refs.~\cite{Amin:2015ftc,Amin:2017wvc,Green:2014xqa}.

\begin{figure}[t]
  \centering
     \includegraphics[width=0.5\textwidth]{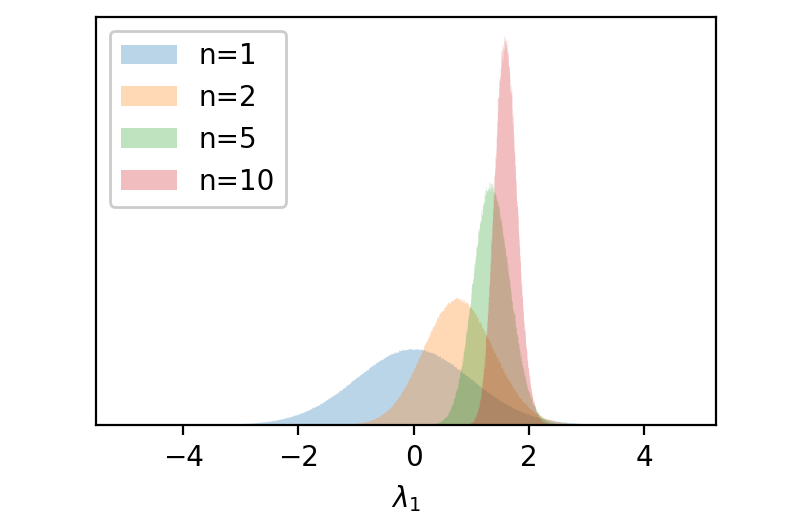}
  \caption{Histograms of the rescaled largest eigenvalue $\lambda/\sqrt{n}$ of $n \times n$ Wigner matrices, with the real and imaginary components of $(M_{n})_{ij}$ drawn from a normal distribution with mean zero and unit variance (this is simply the largest eigenvalue from each random draw used to make the top row of plots in Fig.~\ref{fig:prior_sens_uni}). For large $n$, this distribution is described by the Tracy-Widom distribution.}
  \label{fig:sharp_lambda1}
\end{figure}

Again, a great example of this phenomenon can be found in the study of Wigner matrices. To see this, instead of looking at global statistics of the eigenvalue spectrum, we now focus on a local property: the distribution of the largest eigenvalue of $A_{n}$, which we denote $\lambda_{1}$. As before, we look at Wigner matrices rescaled by $A_{n}/\sqrt{n}$ and with unit variance, $\mathrm{Var}[(A_{n})_{ij}] = 1$.
The cumulative distribution of $\lambda_1$ can be shown to (under general conditions\footnote{This has been proven for broad classes of Wigner matrices but establishing whether or not this is the case for all Wigner matrices remains an open problem \cite{2012arXiv1202.0068T}. The appearance of the Tracy-Widom distribution is common to many diverse complex systems which are said to belong to the KPZ universality class. This phenomenon is poorly understood and is an area of active research.}) converge, for $n \rightarrow \infty$, as
\be\label{eq:TW}
p\left(n^{2/3}(\lambda_{1}-2) \leq t \right) \rightarrow \Xi_{2} \, ,
\ee 
where $ \Xi_{2}$ is the famous Tracy--Widom distribution of the type found in the Gaussian Unitary Ensemble~\cite{Tracy:1992rf}. Indeed the distribution of $\lambda_1$ has a standard deviation that scales as $n^{-2/3}$, and converges almost surely \cite{bai1988necessary} to
\be\label{eq:circle_edge}
\lambda_{1}(A^{n})\rightarrow 2 \, .
\ee
An example of this behaviour is shown in Fig.~\ref{fig:sharp_lambda1} where the distribution of the rescaled $\lambda_1$ of $n\times n$ GUE Wigner matrices is shown for different values of $n$.

With regard to numerically assessing how predictive the model is, in practice we found it simplest to just inspect histograms directly. This is in contrast to exploring sharp transitions and testing for universality, where we found the use of the relative entropy, or mutual information to provide valuable information not so easily appreciated by simply looking at histograms alone.

%

\subsection{Sharp transitions} \label{subsec:sharp}

Sharp transitions in the predictions of a model are ubiquitous in the study of high dimensional probability. A priori, we have no reason to think this will also be true for low-dimensional models. Nevertheless, as discussed in \S\ref{sec:AM_ST}, we do find this behaviour in the axion monodromy model. In this subsection, we again look at the statistics of the largest eigenvalue $\lambda_1$ an example.

\begin{figure}[t]
  \centering
     \includegraphics[width=1.0\textwidth]{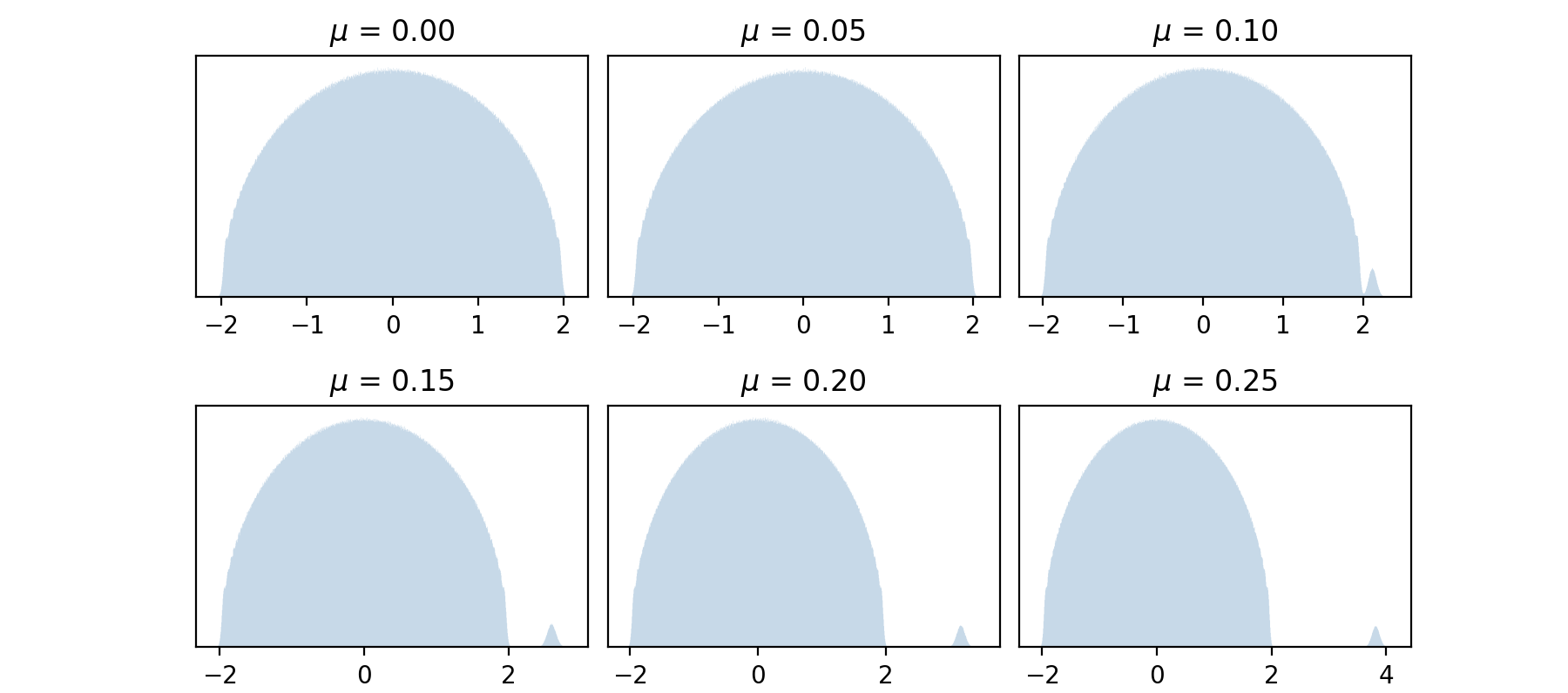}
  \caption{Histograms of the rescaled eigenvalue spectrum of ensembles of Hermitian matrices of the form $A^{n} = (M^{n}+(M^{n})^{\dagger})/2$, with $n = 200$. The real and imaginary components of $(M^{n})_{ij}$ are drawn from a normal distribution with mean $\mu$ and unit variance. When $\mu$ is sufficiently large, the largest eigenvalue gets ejected from the semicircle.}
  \label{fig:mode_sensitivity_bulk}
\end{figure}

For $M_{ij}\sim \mathcal{N}(1,\mu)$, when $\mu=0$ the distribution of $\lambda_1$ is determined by the Tracy-Widom distribution, as described by Eq.~\eqref{eq:TW}.  Once $\mu \neq 0$ the matrix ensemble ceases to be Wigner and the convergence theorems leading to Eq.~\eqref{eq:TW} and Eq.~\eqref{eq:circle_edge} no longer apply. Fig.~\ref{fig:mode_sensitivity_bulk} shows the effect of changing $\mu$ on the eigenvalue spectrum of $A_{n}/\sqrt{n}$. As $\mu$ increases, a second peak appears to the right of the semicircle. Note that this second peak does not violate the result that the spectrum approaches the semicircle law in the large $n$ limit but it does of course violate Eq.~\eqref{eq:circle_edge}, highlighting an important distinction between universality results for local and global properties.

\begin{figure}[t]
  \centering
     \includegraphics[width=0.7\textwidth]{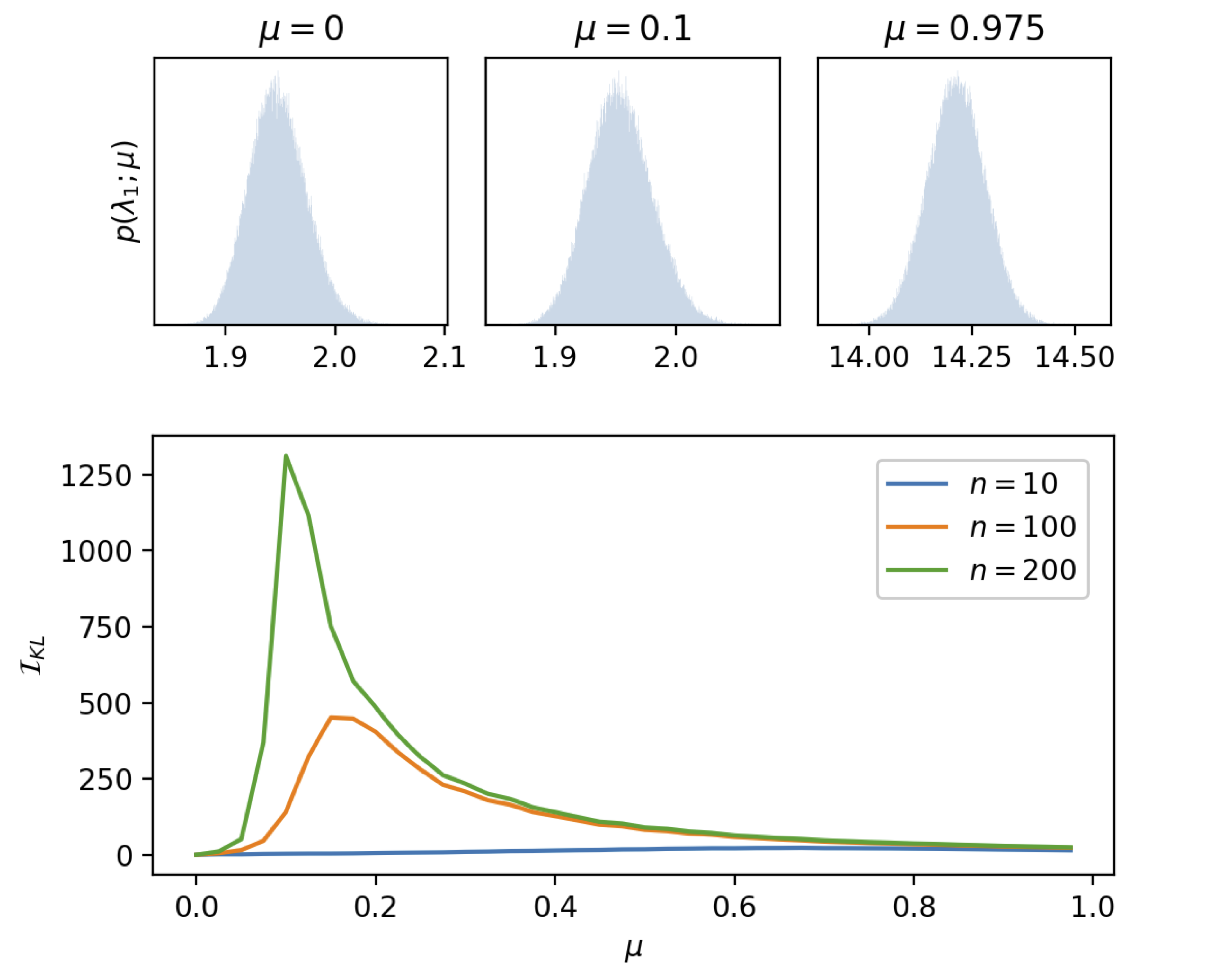}
  \caption{The top row shows histograms of the largest eigenvalue $\lambda_{1}$ of $A^{n}$, with $n = 200$, for different choices of the mean $\mu$ of the prior $(M^{n})_{ij} \sim \mathcal{N}(\mu, 1)$. The lower plot shows $\mathcal{I}_{\rm KL} = D_{\rm KL}(p(\lambda_{1}; \mu = 0)\Vert p(\lambda_{1}; \mu))/D_{\rm KL}(p(M_{ij}; \mu = 0)\Vert p(M_{ij};  \mu))$ for different sizes of Hermitian matrices $n = \{10, 100, 200\}$. As $\mu$ increases, we see a transition in behaviour and the transition becomes sharper with increasing $n$.
 For $\mu = 0$ the distribution closely resembles the Tracy--Widom distribution. For $\mu = 0.975$ the distribution is Gaussian, for $\mu = 0.1$ the distribution neither resembles Tracy-Widom, nor a Gaussian. This is most easily seen by the presence of the sharp spike in $\mathcal{I}_{\rm KL}$ for the $n = 200$ case. }
  \label{fig:D_lambda1}
\end{figure}

To study the distribution of the largest eigenvalue in more detail, we compute
 \be
 \mathcal{I}_{\rm KL} = \frac{D_{\rm KL}(p(\lambda_{1}; \mu = 0)\Vert p(\lambda_{1}; \mu))}{D_{\rm KL}(p(M_{ij}; \mu = 0)\Vert p(M_{ij};  \mu))}\, .
 \ee
 As shown in Fig.~\ref{fig:D_lambda1} the behaviour is quite rich. As $\mu$ increases, the sharp rise in $ \mathcal{I}_{\rm KL}$ tells us that the distribution of $\lambda_{1}$ is initially very sensitive to the distribution of the entries $M_{ij}$, and changes very rapidly. This transition becomes sharper with the size of the matrix (increasing $n$). After the transition, $\mathcal{I}_{\rm KL}$ drops down to a lower value until it settles into a slow rise.
 Fig.~\ref{fig:D_lambda1} also highlights the sensitivity of $\mathcal{I}_{\rm KL}$ to changes in the distribution. The top row of  shows $p(\lambda_{1};\mu)$ for three different choice of $\mu$. The lefthand plot is approximately Tracy--Widom, the righthand plot is well approximated by a Gaussian but the middle plot, at $\mu = 0.1$, according to $\mathcal{I}_{\rm KL}$, differs significantly from either of these. This point is not easily appreciated by looking at the histograms alone. To address this, one could for instance also study the moments of $p(\lambda_{1};\mu)$ but this has clear limitations since in principle there are an infinite number of moments one might need to study and there is no guarantee that any one moment would capture the relevant behaviour. We therefore conclude that $\mathcal{I}_{\rm KL}$ has especially appealing properties as a diagnostic tool. 

Despite not being a member of the Wishart ensemble when $\mu \neq 0$ the behaviour seen in Fig.~\ref{fig:D_lambda1} and Fig.~\ref{fig:mode_sensitivity_bulk} is familiar in random matrix theory. To make contact with known results we decompose $M_{n}$ as
\be
M_{n} =  \tilde{Q}_n + \mathbb{E}[Q_{11}](1+i) J_{n} \, ,
\ee
where $(\tilde{Q}_n)_{ij}$ are iid with zero mean and $J$ is the all-ones matrix. $A_{n}$ then takes the form
\be
A_{n} = W_{n} + \mathbb{E}[Q_{11}]J_{n} \, ,
\ee
where $W_{n}= \frac{1}{2}\left (\tilde{Q}_n + (\tilde{Q}_n)^\dagger \right )$ is a Wigner matrix and the second term is a deterministic rank one matrix. Thus, varying the mean $\mu$ is a particular case of finite rank deformations of Wigner matrices, for which there are a number of known results. In particular, Refs.~\cite{2007arXiv0706.0136C, 2007CMaPh.272..185F} describe in detail very similar behaviour to that found here. Interestingly, Ref.~\cite{2007arXiv0706.0136C} points out that the limiting distribution of $\lambda_{1}$ depends on the form of the deformation\footnote{A particularly interesting form of deformation is a diagonal rank one matrix. In this case Ref.~\cite{2007arXiv0706.0136C} find non-universality and the distribution of $\lambda_1$ is a convolution between a Gaussian and the distribution of the entries of $W_{n}$.}. For the case of interest here, where the deformation is proportional to $J_{n}$, they find universal behaviour and the distribution of $\lambda_{1}$ is Gaussian for large deformations.

\subsection{Overlaps with other fields of research}\label{overlaps} 
The usage of information theory to study parameter sensitivity is not new. For instance, in the context of artificial neural networks, this falls under the broader subject of input variable selection methods~\cite{may2011review}. Recently, the Fisher information matrix (FIM) was used to study Higgs measurements at the LHC~\cite{Brehmer:2016nyr} and similar in spirit to this is the body of work on so-called `sloppy' models such as in Refs.~\cite{brown2003statistical, brown2004statistical, waterfall2006sloppy}. In these cases they use hierarchies between in the eigenvalues of the FIM to identify hierarchies in parameter sensitivity. This is rather similar to our usage of the relative entropy, in so much as in information geometry, the FIM can be viewed as a metric in the same sense that the relative entropy plays the role of a distance measure (see \emph{e.g.} Ref.~\cite{ay2017information} for a review). We explored the possibility of using the FIM in this work but found the usage of relative entropy to have a number of practical advantages. Apart from the various pros and cons of using local \emph{vs} non-local quantities as diagnostic tools, we also found the relative entropy considerably easier to compute.

Finally, in the context of Bayesian inference there is the notion of ‘model
complexity’, or ‘Bayesian complexity’ \cite{spiegelhalter2002bayesian, kunz2006measuring} which seeks to compute the effective
number of parameters of a model which can be inferred from data.


\section{Our test model: Axion Monodromy} \label{sec:model}

To test the idea of robust predictions and handling of theoretical uncertainties in a specific inflationary model we turn to axion monodromy. This model, as we will discuss in the next section, works as a good implementation example as it is specific enough to have UV input for the shape of the potential but also accommodates theoretical uncertainty as suppressed corrections to the tree level potential.

\subsection{A brief review of axion monodromy}

\subsubsection{Historical progression of axion inflation}

Inflation is a process whose effective field theory (EFT) description is sensitive to the UV completion of its underlying dynamical mechanism in a putative theory of quantum gravity. In the context of canonical single-field slow-roll inflation driven by a scalar field $\phi$ with scalar potential $V(\phi)$ this sensitivity shows up in the form of an infinite series of higher-dimension operators
\be
\delta V_{4+n}=V(\phi)\frac{\phi^n}{\Mp^n}\quad.
\ee
They induce corrections
\be
\delta\epsilon_n \sim \left(\frac{\phi^{n-1}}{\Mp^{n-1}}\right)^2\quad,\quad \delta\eta_n \sim \frac{\phi^{n-2}}{\Mp^{n-2}}
\ee
to the slow-roll parameters $\epsilon$, $\eta$ which become $\MC{O}(1)$ for all $n\geq 1$ and $n\geq 2$, respectively, once $\phi\gtrsim \Mp$.
However, assuming monotonicity of $\epsilon$ along the the slow-roll trajectory, the Lyth bound~\cite{Lyth:1996im}
\be\label{eq:Lyth}
r\lesssim 8 \left(\frac{\Delta\phi_{N_e}}{N_e}\right)^2\simeq 0.003\times \left(\frac{\Delta\phi_{N_e}}{\Mp}\right)^2 \left(\frac{50}{N_e}\right)^2
\ee
on the tensor-to-scalar ratio $r\equiv P_\zeta/P_\MC{T}=16\epsilon$ endows those single-field models which traverse a field range $\Delta\phi_{60}\gtrsim \Mp$ during the last observable 60-something e-folds of inflation with detectable amounts of primordial tensor modes. Hence, it is exactly this most UV sensitive class of `large-field inflation' which at the same time offers enhanced testability by producing detectable levels of $r$.

Clearly, calculational control of such `large-field inflation models' requires suppression of an infinite series of higher-dimension operators in the effective action. This amounts to requiring that UV completions of large-field inflation must provide a sufficiently softly broken effective shift symmetry for the inflaton $\phi$.

For these reasons, we try to construct inflation models in controlled compactifications of string theory as a candidate theory of quantum gravity. At first, it was realised that most of the volume moduli of the extra dimensions, and the D-brane position moduli used in the first string inflation models, generically have compact moduli spaces with diameter $\lesssim \Mp$. In consequence,  controlled large-field inflation models of the type $V\sim \phi^p$ are built in string theory so far utilising the discrete shift symmetries of the axions which descend from the higher-dimensional $p$-form gauge potentials of the theory.
The shift symmetries of these string axions are only broken by non-perturbative quantum effects, D-branes or fluxes, leading to expectations of enhanced control over the infinite series of dangerous corrections discussed before.

However, a wealth of combined theoretical evidence both from the direct analysis of different classes of string compactifications~\cite{Banks:2003sx,Svrcek:2006yi} as well as low-energy effective quantum gravity constraints such as the weak gravity conjecture (WGC)~\cite{ArkaniHamed:2006dz} has established a constraint on the axion decay constants $f_i\lesssim \Mp$ as an `empirical' fact. Concretely, axions $c_n^i$ descending from $n$-form gauge potentials $C_n$ in ten dimensions lead in four-dimensional compactifications controlled by a single length scale $L=\MC{V}^{1/6}$ to axion decay constants
\be
f_{c_n^i}\sim \frac{\Mp}{L^{2n}}=\frac{\Mp}{\vol^{n/3}}\quad.
\ee
Here, $\vol=L^6$ denotes the dimensionless volume of the compact six extradimensions in units of the string scale $\ell_s=\sqrt{\alpha'}$. We can arrange for a setup, where we localise the $n$-dimensional submanifolds, called $n$-cycles $\Sigma_n^i$, on which the $C_n$-axions $c_n^i$ and their kinetic terms live, in a warped region of the extra dimensions.  The axion decay constants arise from canonically normalising the kinetic terms of the axions.  Hence, if an axion kinetic term is localised in a warped region of the extra dimensions, then the warp factor $\exp(A)$ at the location of the axion kinetic term (the IR redshift of the warped region) suppresses the decay constants 
\be\label{eq:DecayConst}
f_i\equiv f_{c_n^i}\sim \Mp \frac{e^{A}}{\vol^{l/6}}
\ee
where we define $l\equiv 2n \in \{2,4,6\}$. This again implies $f_i< \Mp$ as by definition 
\be\label{eq:warping}
\exp(A)<1
\ee
in a warped region, and 
\be\label{eq:SugraCrit}
\vol > 1
\ee 
for computational control of the supergravity approximation to the full string theory. As the fundamental region of periodicity of the axions $2\pi f_i$ defines their field range, this seemed to preclude large-field inflation with axions in string theory.

\subsubsection{Basics of axion monodromy}

This conclusion turned out to be too na\"ive due to two realisations. Firstly, alignment effects in the effective action of a few axions~\cite{Kim:2004rp}, as well as the collective Hubble friction assistance of the many axions of N-flation~\cite{Liddle:1998jc, Copeland:1999cs, Dimopoulos:2005ac} provide examples where the collaboration of several axions increase the total axion field range by providing a covering space for one effective axion direction which covers many fundamental domains. Secondly, restricting from now on to the case of a single axion, the analysis of Refs.~\cite{Silverstein:2008sg,McAllister:2008hb,Kaloper:2008fb} exhibited the existence of sources of axion stress-energy which intrinsically break the discrete shift symmetry of an axion: these sources display `monodromy' under a shift $\phi\to \phi+2\pi f_\phi$ of the canonically normalised axion $\phi$ arising from a spontaneous breaking (`Higgsing') of the 10D $U(1)$ symmetry underlying the discrete axion shift symmetry. For a discussion of large-field inflation in string theory, see \emph{e.g.} Ref.~\cite{Baumann:2014nda} for the definitive textbook and \emph{e.g.}~\cite{Cicoli:2011zz,Westphal:2014ana} for some reviews.
Such sources of stress-energy with axion monodromy can be \emph{e.g.} D-branes or NS5-branes coupled to the higher $p$-form gauge potentials from which the axions descend, axion-$p$-form flux Chern Simons couplings, or $p$-form field strengths themselves~\cite{McAllister:2008hb,Dong:2010in,Palti:2014kza,Marchesano:2014mla,Hebecker:2014eua,McAllister:2014mpa,Ibanez:2014swa}.

Subsequent work~\cite{Dong:2010in,Kaloper:2011jz,Kaloper:2014zba,McAllister:2014mpa,Hebecker:2014kva,Buchmuller:2015oma,Landete:2016cix,Bielleman:2016grv,Valenzuela:2016yny,Bielleman:2016olv,Baume:2016psm,Blumenhagen:2017cxt,Landete:2017amp,Blumenhagen:2018hsh} led to a more differentiated picture where the presence of the moduli fields of string compactifications as well as certain non-perturbative quantum corrections leads to backreaction effects on the string axion inflaton potential. These backreaction effects can sometimes break the underlying discrete shift symmetry badly enough to destroy inflation~\cite{Hebecker:2014kva,Buchmuller:2015oma}. In other cases~\cite{Dong:2010in,Kaloper:2011jz,Kaloper:2014zba,McAllister:2014mpa,Landete:2016cix,Bielleman:2016grv,Bielleman:2016olv,Landete:2017amp} these effects respect the underlying periodicities of the discrete shift symmetries and produce the generic effect of \emph{flattening} the axion inflation potential 
\be\label{eq:FlattenedPot}
V^{(0)}(\phi)\sim \phi^{p_0}\quad\to\quad V_{\rm tree}(\phi)=V_0\left[\left(1+\frac{\phi^2}{\mu^2}\right)^{p/2} -1 \right]\sim \phi^p\quad,\quad \phi\gg \mu
\ee
such that generically $p<p_0$. Existing lamp post constructions~\cite{Silverstein:2008sg,McAllister:2008hb,Dong:2010in,McAllister:2014mpa,Landete:2017amp} with $p_0=2$ have produced $p\in\{2/5,2/3,1,4/3\}$ while the cross-over field range $\mu$, which is typically given by a combination of the volume of $n$-cycle carrying the $c_n$-axion and the decay constant $f_\phi$, has led to $\mu\lesssim \Mp$. Hence, we will describe $p$ as being drawn from a prior distribution with range $p\in\{0.1,2\}$ and varying shape, while $\mu$ will be described as being drawn from a prior with range $p\in\{0.1,1\}$ and varying shape.

Moreover, we find from these lamp-post constructions of axion monodromy, that the overall scale of the inflation potential $V_0$ arises both from the Weyl rescaling from 10D string frame to 4D Einstein frame providing a factor $1/\vol^2$ and potential redshifting $e^{4A}$ from placing the source of axion monodromy at the infrared end of warped throat regions of the extra dimensions as \emph{e.g.} in Ref.~\cite{Kachru:2003aw,McAllister:2008hb}
\be\label{eq:PotScale}
V_0 \sim \frac{e^{4A}}{\vol^2}\quad.
\ee
The axion monodromy induced potential for the axion is subject to the warp factor suppression, if \emph{e.g.} the monodromy-generating D-branes or fluxes providing the axion potential are localised at the IR end of such a warped throat.

The warp factor of such throat regions provides for the possibility of exponentially small scales. However, consistent string compactifications realising such warping in type IIB string theory~\cite{Giddings:2001yu} require the presence of quantised 3-form fluxes. Each such warped throat region of a CY manifold arises from a complex structure modulus being stabilised by 3-form flux at exponentially small values. This requires at least one pair of quantised 3-form fluxes with quanta $Q$ and $M$ on the $A$ and $B$ 3-cycle, respectively, of a given warped throat. In terms of these, the warp factor at the IR end of the throat becomes~\cite{Giddings:2001yu}
\be\label{eq:WarpFlux}
e^{A}\sim e^{-\frac{Q}{M}}\quad.
\ee 

An important constraint on the distribution of $Q$ and $M$ comes from the fact that the 3-form fluxes contribute to the total amount of D3-brane charge $Q_{D3}$. The compactness of the extra dimensions then dictates a Gauss law type `tadpole' constraint limiting the maximum value of $Q_{D3}$. As the pair of quantised 3-form fluxes with quanta $Q$ and $M$ in a given warped throat contributes an amount $Q\cdot M$ to $Q_{D3}$, we see that we must impose on any model of axion monodromy a limit 
\be\label{eq:D3tadpole}
Q\cdot M < Q_{D3}<Q_{D3}^{max}\quad.
\ee
We will thus describe $Q$ and $M$ as being drawn distributions with support on
\be\label{eq:QMprior}
Q,M\in [1,Q_{D3}^{max}]
\ee
subject to the above tadpole constraint. The maximum possible D3-brane charge in such type IIB string compactifications we estimate by looking at the F-theory lift of the type IIB string compactification with the largest Euler number of the underlying elliptically fibred CY fourfold which determines $Q_{D3}^{max}$ to be $Q_{D3}^{max}\sim 10^5$~\cite{Taylor:2015xtz}.

\subsubsection{Corrections to the tree-level model}
Beyond the dominant flattening backreaction effects, the underlying discrete shift symmetries controlling many of the corrections to such axion monodromy inflation models show up in the fact that perturbative higher-dimensional corrections in 10D typically arise as integer powers of the field strengths $|F_p|^2$ or their couplings to powers of the Riemann curvature tensor, or nonperturbative effects producing periodically oscillating corrections.  Hence, many of the corrections take the forms
\be\label{eq:Corrections1}
V_{\rm p}^{\rm eff.}= \sum^N_{n \geqslant 2} \chi_n \  C_{n}^{\rm eff.} \frac{\left(V^{(0)}(\phi)\right)^n}{M_{\rm p}^{4n -4}}\quad,\quad V_{\rm np}=\Lambda^4_{\rm UV}\sum^M_{m \geqslant 1} D_m e^{-m S} \cos{\left(\frac{m\phi}{f_\phi}\right)}
\ee
where generically the instanton actions $S$ appearing in the non-perturbative effects may have dependence on the inflaton-axion $\phi$ as well. Moreover, $V^{(0)}(\phi)=m^2\phi^2$ denotes the tree-level axion scalar potential arising from the quadratic $|F_p|^2$-terms in 10D. The instanton action for an instanton wrapping a $k$-cycle of the extra dimensions will scale as
\begin{equation}\label{eq:InstAction}
S= \mathcal{C}\, \vol^k
\end{equation}
for dimensional reasons, with $k \in \{2,3,4,6\}$ and $\mathcal{C} \in [\vol^{-l},1]$. 
The energy scale $\Lambda_{UV}$ of the non-perturbative effects, by Weyl rescaling from 10D string frame to 4D Einstein frame, behaves as 
\begin{equation}\label{eq:lambdaUV}
\Lambda_{UV}^4\sim\frac{1}{\vol^2} \quad .
\end{equation}
This is typically to be expected, because non-perturbative effects arise in string theory from branes wrapping cycles of the compact extra-dimensions. Hence, the Kaluza-Klein scale of the extra-dimensions generically provides the UV cut-off to these non-perturbative corrections which in turn leads to the above behaviour of $\Lambda_{UV}^4$.

\begin{figure}[t]
  \centering
     \includegraphics[width=0.53\textwidth]{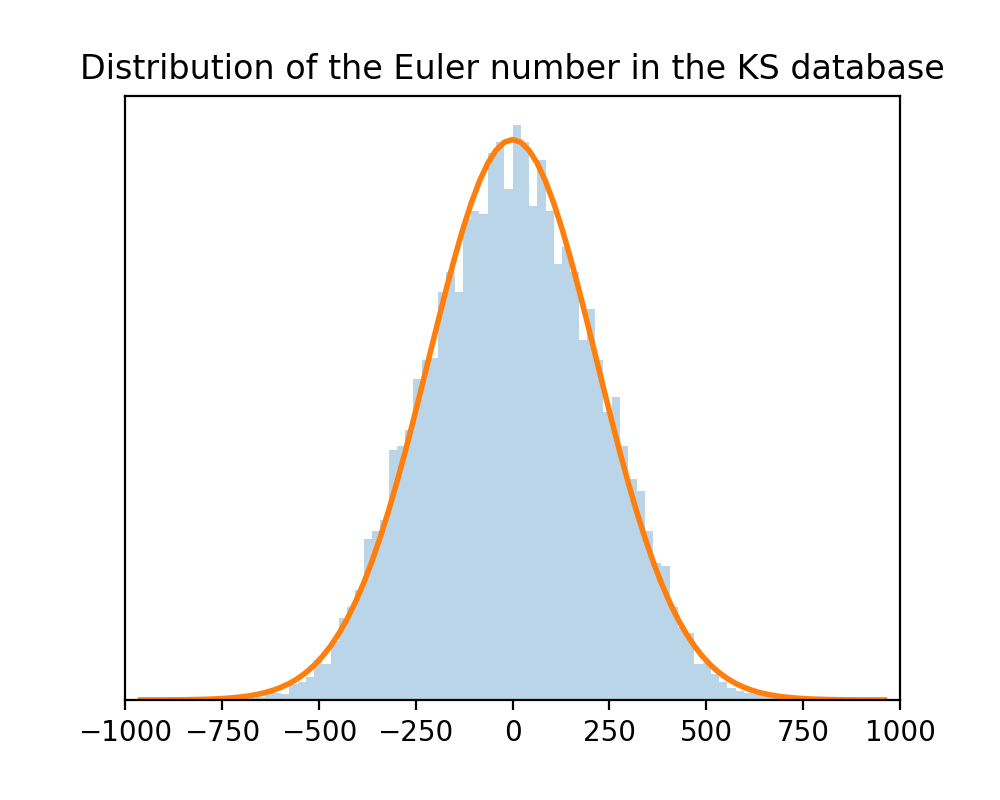}
\caption{Histogram of the 2nd and 3rd Chern classes (Euler number) in the Kreuzer-Skarke database~\cite{Kreuzer:2000xy}, together with a Gaussian PDF with mean zero and standard deviation of 217.}
  \label{fig:euler}
\end{figure}

Moving on, the $C_n$ and $D_m$ denote generically $\MC{O}(1)$ Wilsonian EFT coefficients, while the $\chi_n$ represent the fact that many of the perturbative higher-dimensional corrections in 10D typically arising as integer powers of the field strengths $|F_p|^2$ or their couplings to powers of the Riemann curvature tensor are controlled by topological invariants of the extra dimensional manifold. We collectively denote such invariants with $\chi_n$ and use the example of the 2nd and 3rd Chern classes (Euler number) as a guide for the typical sizes: \emph{e.g.} for the Kreuzer-Skarke database of CY manifolds~\cite{Kreuzer:2000xy} the typical size of the Euler number is $\simeq 200$, see Fig.~\ref{fig:euler}.

Next, we note that the non-perturbative corrections in Eq.~\eqref{eq:Corrections1} generate oscillating terms in the scalar potential. These oscillation patterns cause oscillating features in the curvature perturbation two-point function power spectrum, and enhanced levels of non-Gaussianity from resonance effects~\cite{Flauger:2009ab}. Treating these interactions within the effective field theory of inflation~\cite{Cheung:2007st}, described by just the axion field fluctuation coupled to the Goldstone boson of dS time translation invariance breaking, places a lower bound of
\be\label{eq:fLower}
f_\phi\gtrsim 10^{-4}\Mp
\ee
on the axion decay constant. Otherwise, the effective theory would be forced to include new degrees of freedom. These get excited once $f_\phi<10^{-4}\Mp$ because then the oscillating interactions for the slow-rolling axion-inflaton have a time-domain frequency above the UV cutoff of the EFT of inflation $\omega>\Lambda$~\cite{Flauger:2014ana}.

In parallel, a series of works~\cite{Kaloper:2008fb,Kaloper:2011jz,Kaloper:2014zba,Kaloper:2016fbr} established that the 4D effective description of axion monxinflation can be cast in the form of an effective 4-form field strength $F_4=dA_3$ coupled to the inflaton-axion $\phi$ by a CS-term
\be
\Delta\MC{L}=\phi\epsilon_{\mu\nu\rho\sigma}F^{\mu\nu\rho\sigma}\quad.
\ee
Combined with the kinetic term for $F_4$ which together with Lorentz invariance forces $F_{\mu\nu\rho\sigma}= const. \times \epsilon_{\mu\nu\rho\sigma}$, this generates a quadratic potential for $\phi$ by integrating out $F_4$. As this happens while the full action for $\phi$ and $F_4$ possesses the full combined gauge and shift symmetries of $F_4$ and $\phi$, this shows monodromy as a spontaneous symmetry breaking phenomenon.

Subsequent works~\cite{Kaloper:2011jz,DAmico:2017cda} demonstrated that higher-dimension corrections respecting the underlying symmetries of the 4D effective action must come essentially as integer powers of $F_4^2$ or non-perturbative corrections. As the tree level potential of $\phi$, $V^{(0)}=m^2\phi^2$ arises from the tree-level $F_4^2$ term and the $\phi F_4$ coupling, this shows that the corrections come in powers of $V^{(0)}$ which nicely matches with the structure of the corrections descending from 10D in the string models above.

From this discussion it is clear that essentially all of the corrections consistent with 4D symmetries of the Kaloper-Sorbo effective description of axion monodromy take the form of Eq.~\eqref{eq:Corrections1}. Moreover, the fact that these corrections arise in the Kaloper-Sorbo descriptions as terms $(F_4^2)^n$, implies that the perturbative corrections of the type  Eq.~\eqref{eq:Corrections1} can be decomposed for each $n$ into two contributions
\be
\chi_n C_n^{\rm eff.}\equiv C_n^{\rm tree}+\chi_n C_n\quad.
\ee
Here, we model the parameter $ C_n$ as a random $\MC{O}(1)$ Wilsonian EFT coefficient, while we split off a contribution $C_n^{\rm tree}$ such that all the associated correction pieces
\be\label{eq:V_tree}
V_{\rm p}^{\rm tree}= \sum^N_{n \geqslant 2}  C_{n}^{\rm tree} \frac{\left(V^{(0)}(\phi)\right)^n}{M_{\rm p}^{4n -4}}=V_0\left[\left(1+\frac{\phi^2}{\mu^2}\right)^{p/2} -1 \right]\quad,\quad V_{\rm p}=\sum^N_{n \geqslant 2}  \chi_n  C_{n} \frac{\left(V^{(0)}(\phi)\right)^n}{M_{\rm p}^{4n -4}}
\ee
resum into the flattened tree-level axion inflaton potential $V_{\rm tree}(\phi)$.

Hence, we expect that describing single-field axion monodromy inflation as arising from a flattened tree-level potential Eq.~\eqref{eq:FlattenedPot} subject to an infinite series of corrections Eq.~\eqref{eq:Corrections1} exhausting the range allowed by 4D symmetries underlying the 4D Kaloper-Sorbo effective description of axion monodromy, should capture a large class of effects arising from the typical spectrum of corrections we expect in a string theory model of axion monodromy inflation. As such, this setup forms the basis for a probabilistic network analysis of single-field axion monodromy inflation to which we now turn.

\subsection{Introducing the probabilistic model}\label{sec:PGM_tree}

Following the discussion of the previous section, the model we will be working with is an axion monodromy construction based on Ref.~\cite{Dong:2010in} where suppressed perturbative and non-perturbative corrections to the potential are modelled in the spirit of Ref.~\cite{Kaloper:2011jz}. 
The potential looks like 
\begin{equation}
\label{eq:full_potential}
V=V_{\rm p}^{\rm eff.}+V_{\rm np}=\underbrace{V_0\left[\left(1+\frac{\phi^2}{\mu^2}\right)^{p/2} -1 \right]}_{V_{\rm p}^{\rm tree}} + V_{\rm p} + V_{\rm np} \ ,
\end{equation}
where $V_{\rm p}$ and $V_{\rm np}$ are defined by Eqs.~\eqref{eq:Corrections1} and \eqref{eq:V_tree}.

\begin{figure}[t]
  \centering
     \includegraphics[width=0.8\textwidth]{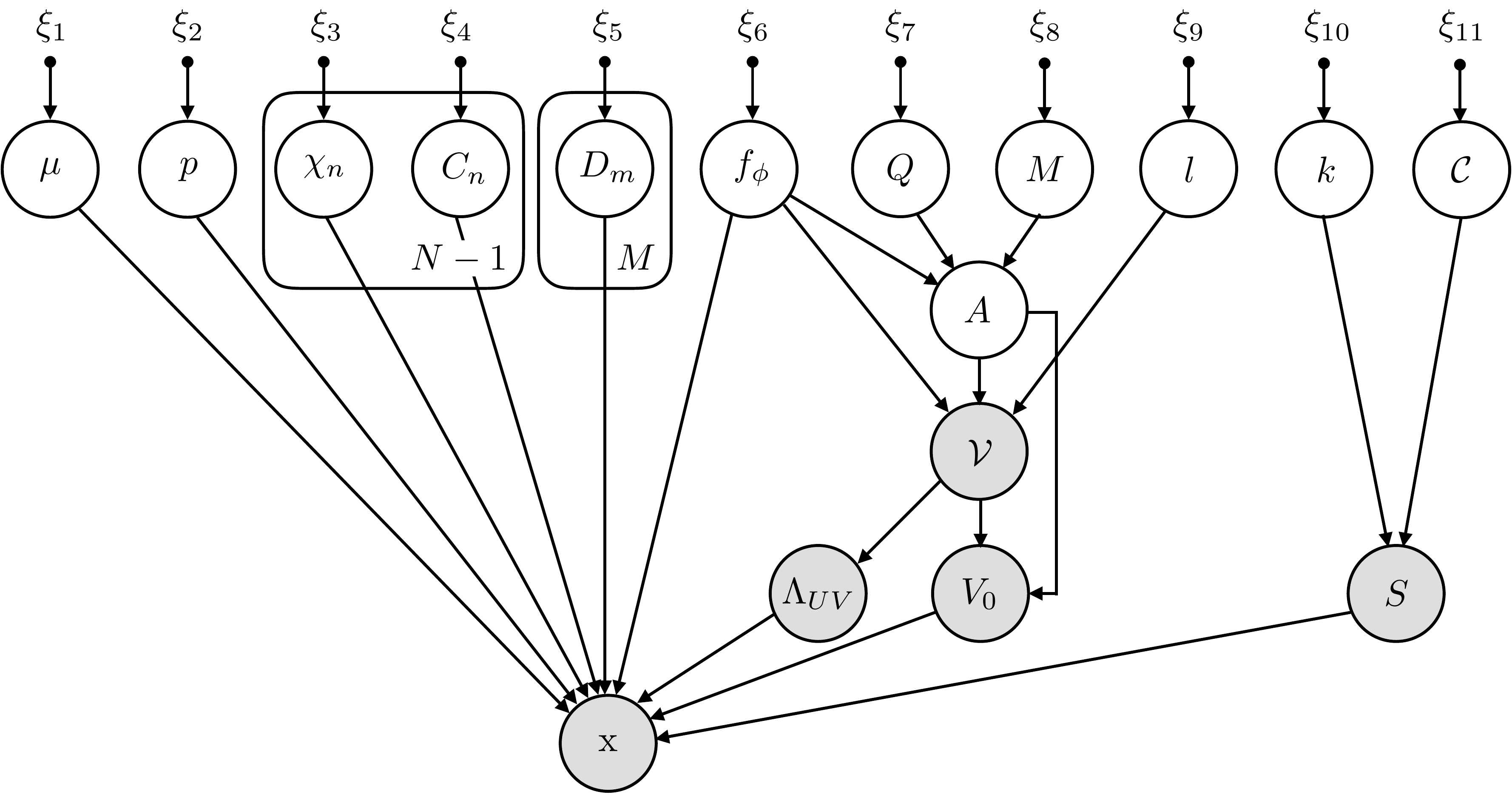}
  \caption{Graphical representation of the full potential of Eq.~\eqref{eq:full_potential}}
  \label{fig:graph_total}
\end{figure}

As discussed in \S\ref{sec:PGM}, the first step to studying this model systematically is to recast it as a probabilistic graphical model where all dependencies are made clear, and to establish the relevant range of values each parameter can take. Our proposed graphical representation of the model is presented in Fig.~\ref{fig:graph_total}. We can see that the parameters in the potential do not necessarily correspond to the parameters which are most easily constrained from the UV perspective. Rather, in some cases, they are related by a series stochastic and deterministic dependencies. The deterministic dependencies have been given in the previous section in Eqs.~\eqref{eq:DecayConst}, \eqref{eq:lambdaUV}, \eqref{eq:PotScale} and \eqref{eq:InstAction}, together with the expressions for observables given by Eqs.~\eqref{eq:Pz}, \eqref{eq:ns}, \eqref{eq:as} and \eqref{eq:r}. We define the stochastic dependencies in table \ref{table:parameters}  by giving the range of scales of each stochastic variable and their fiducial prior. In cases where the top-level parameters have a finite range, we assume a (log-) uniform distribution. This is simply a statement of our ignorance (\cite{laplace2012pierre}) of further microphysical considerations. In some other cases we have a little more information. Most notably, our choice of fiducial prior for $\chi_n$ comes from looking at the distribution of the 2nd and 3rd Chern classes (Euler number) in the Kreuzer-Skarke database~\cite{Kreuzer:2000xy}. Fig.~\ref{fig:euler} shows a histogram of this data, together with a Gaussian PDF with zero mean and standard deviation of 217. This very crude matching will prove to be more than sufficient for our needs in \S\ref{sec:results_full}.

Ultimately however the precise choice of fiducial prior is of limited importance. To probe the model's sensitivity to the choice of prior we endow the top-level priors with hyperparameters $\xi$ which enable parametric deviations from the fiducial prior. We must therefore choose a distribution which has finite range, includes the uniform distribution for a special choice of the hyperparameters, and yet also demonstrates sufficient flexibility to enable us to study the impact of significant variations of the prior. We find the Beta distribution to meet these requirements quite nicely. 

\begin{figure}[t]
  \centering
     \includegraphics[width=0.95\textwidth]{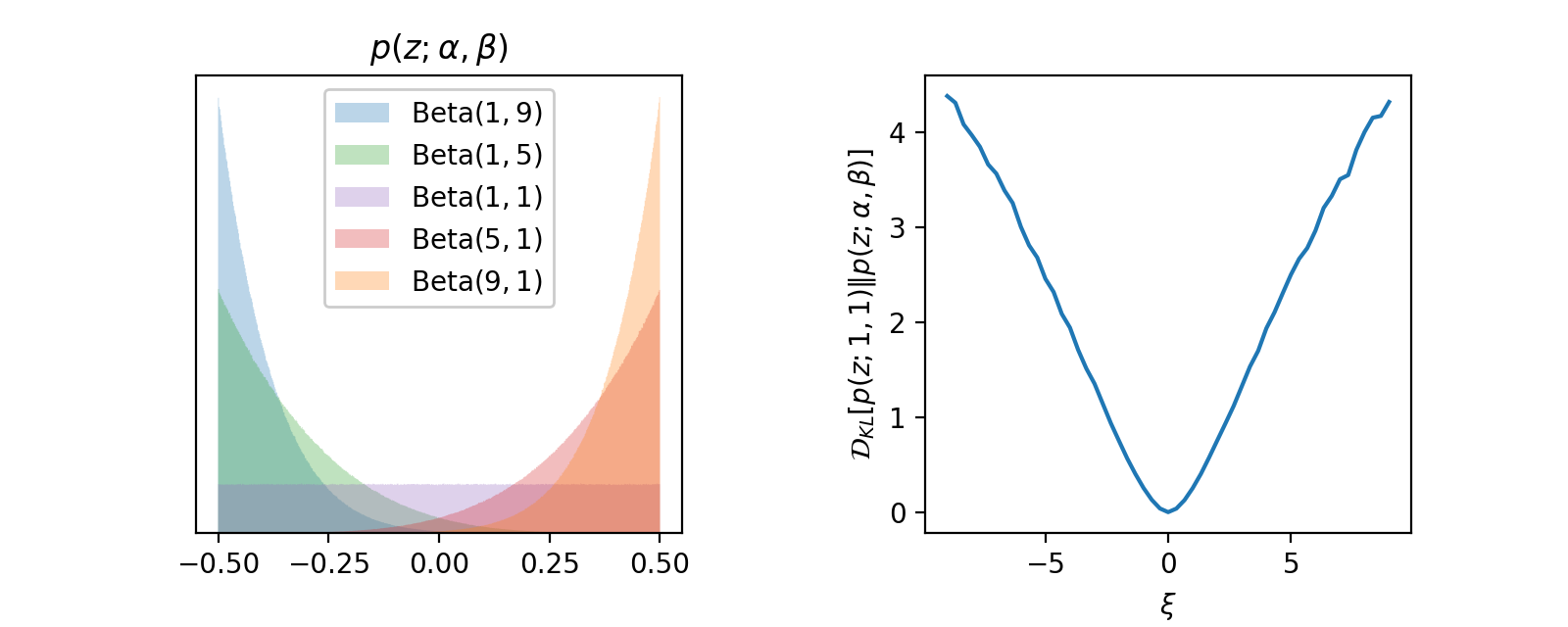}
\caption{Beta distribution: On the left hand plot, the distribution is shown for different values of $\alpha$ and $\beta$. The right hand plot shows the increase in KL-divergence (relative entropy) between the fiducial prior ($\alpha=1$, $\beta=1$) and the Beta distribution as $\alpha$ and $\beta$ are varied. The parameter $\xi$ is positive for fixed $\beta=1$ such that $\xi=\alpha$, and it is negative for fixed $\alpha=1$ such that $\xi=-\beta$.}
  \label{fig:Beta}
\end{figure}

The Beta distribution has the probability density function
\be
{\rm p}(z; \alpha, \beta) = \frac{\Gamma(\alpha+\beta)}{\Gamma(\alpha)\Gamma(\beta)} z^{\alpha - 1}(1-z)^{\beta - 1} - \frac{1}{2} \, .
\ee
It has support on the interval $[-1/2,1/2]$, which we rescale for each parameter to match the desired range and for $\alpha = 1$ and $\beta = 1$ is equal to the uniform distribution. By varying the hyperparameters $\alpha$ and $\beta$, we can obtain a wide variety of distributions. In practice though, we do not perform a complete scan of both $\alpha$ and $\beta$. Instead we define a new hyperparamter $\xi$ which for positive values is gives $\xi = \beta$ and $\alpha = 1$ and for negative values $\xi = -\alpha$ and $\beta = 1$. Examples of the resulting distribution are shown in the lefthand plot of Fig.~\ref{fig:Beta}. The righthand plot shows how the relative entropy between the fiducial distribution ($\alpha = 1$, $\beta  = 1$) and the Beta distribution depends on $\xi$.

For the parameters with Gaussian fiducial prior, we simply study the impact of varying the mean. For all priors, we compare the same range of values of the relative entropy and we choose the maximum relative entropy to be roughly 4.4. So for instance, if a given parameter were to have a fiducial prior that is Gaussian with mean zero and 0.5 standard deviation, then the maximum variation of the mean would be 1.5, such that the maximum relative entropy between the fiducial prior and the prior with shifted mean would be roughly 4.4. Similarly for the Beta distribution, we take the maximum values of $\xi$ to to be 9, since, as shown in Fig.~\ref{fig:Beta}, this gives a relative entropy with respect to the uniform distribution of roughly 4.4.

\begin{table}[h]
\caption{Stochastic variables of the model described by Fig.\ref{fig:graph_total}}
\begin{tabular}{@{}p{0.13\textwidth}*{4}{L{\dimexpr0.28\textwidth-2\tabcolsep\relax}}@{}}
\toprule
    & {range of scales} & {fiducial prior} & {relevant discussion}  \\ 
 \midrule
    {$\mu$}  &  [0.1 , 1] &  flat distribution, by lack of particular structure & around Eq.~\eqref{eq:FlattenedPot}   \\
    \midrule
    {$p$} & [0.1 , 2]  & flat distribution, by lack of particular structure & around Eq.~\eqref{eq:FlattenedPot}    \\
    \midrule
    {$\chi_n$} & $\mathcal{O}(200)$  & ${\mathcal N}(0,200)$ & after Eq.~\eqref{eq:InstAction}   \\
    \midrule
    {$C_n$} & $\mathcal{O}(1)$  &  ${\mathcal N}(0,1)$ & after Eq.~\eqref{eq:InstAction}   \\
    \midrule
    {$D_m$} & $\mathcal{O}(1)$  &  ${\mathcal N}(0,1)$ & after Eq.~\eqref{eq:InstAction} \\ 
    \midrule
    {$f_\phi$} &  $\left[10^{-4}, 1\right]$  &  log-flat prior, by lack of particular structure & WGC arguments together with discussion around Eq.~\eqref{eq:fLower}\\ 
    \midrule
    {$e^A$} & $\left[f_{\phi}, 1\right]$  & distribution defined by $Q$ and $M$ according to Eq.~\eqref{eq:WarpFlux} & the range of scales is determined by Eq.~\eqref{eq:DecayConst}, together with the prior on $f_{\phi}$ and Eq.~\eqref{eq:warping} and Eq.~\eqref{eq:SugraCrit}\\ 
    \midrule
    {$Q$ and $M$} & $\left[1, 10^{5}\right]$ conditioned on $Q\cdot M<10^{5}$ & flat distribution, by lack of particular structure & around Eqs.~\eqref{eq:D3tadpole} and \eqref{eq:QMprior}  \\ 
    \midrule
    {$l$} & $\{2,4,6\}$   & flat  &  see Eq.~\eqref{eq:DecayConst} \\ 
    \midrule
    {$k$} & $\{2,3,4,6\}$  & flat & around Eq.~\eqref{eq:InstAction} \\ 
    \midrule
    {$C$} &  $[\vol^{-l},1]$ &  flat distribution, by lack of particular structure & around Eq.~\eqref{eq:InstAction} \\
\bottomrule
\end{tabular}
\label{table:parameters}
\end{table}

Beside the parameters of the potential,  the model in principle also depends on the initial conditions for the field and its velocity, $\phi_0$ and $\dot\phi_0$, We chose not to model this dependence, by requiring 55 e-folds of inflation from all realisations. Similarly, we also do not model the choice of pivot scale at which the observables are estimated (more specifically, the number e-folds from the end of inflation to horizon exit of the pivot scale). The uncertainty in the pivot scale is related to the uncertainty in the reheating process. For simplicity, we assume instantaneous reheating. 

Before moving on to the analysis of this model, we would like to make a comment on the range of $V_0$. According to Eq.~\eqref{eq:PotScale}, and looking at the range of scales of $A$ and $\vol$, the scale $V_0$ can take values between $\left[10^{-24} , 1\right]$. This is problematic as $V_0 \sim 1$ breaks the perturbative treatment of this potential. For our purposes, we will condition $V_0 < 10^{-5}$, which ensures that the potential is self-consistent while keeping the range of values broad enough to safely capture all interesting physics compatible with our observable universe. It will become clearer in what follows that looking at larger values of $V_0$ is indeed of little physical interest.

\section{Warmup: Analysis of the Tree-Level Axion Monodromy Model}\label{sec:results_tree}
The tree-level model, illustrated in Fig.~\ref{fig:AM_graph_simple}, is especially straightforward to study. Since slow-roll conditions hold, the only observable which depends on $V_{0}$ is the amplitude of the power spectrum $P_{\zeta}$, while the other observables $n_s$, $\alpha_s$ and $r$ only depend on two parameters, $\mu$ and $p$. We therefore treat this model as a warmup to help gain some intuition before moving on to the full model in the next section.

\subsection{Machine learning the mapping from model parameters to observables}
Our analysis relies heavily on being able to repeatedly generate large samples of observables for different choices of priors. As a first step, for the fiducial prior, we follow the strategy below:
\begin{itemize}
\item Take a random draw of the model parameters (for the tree level model this is simply $\{V_{0}, \mu, p\}$)
\item Numerically solve the equations of motion for the background and perturbations, starting with initial conditions that ensure at least 60 $e$-folds of inflation
\item Compute $\{P_{\zeta}, n_s, \alpha_s,r\}$
\item Repeat this procedure, say, $\mathcal{O}(10^7)$ times for a fixed choice of hyperparameters
\end{itemize}
We use a minimally adapted version of the Transport code \cite{Dias:2015rca, Dias:2016rjq} to numerically solve the background equations of motion and compute observables. In principle we could perform the above procedure for each choice of prior that we wish to study. However, this becomes computationally quite costly and certainly highly inefficient, since we would be repeatedly sampling the same parameter space, just with a different weighting each time we change the prior. Instead we choose to ``learn'' the mapping from model parameters to observables. Specifically, for the tree level model we wish to find numerical approximations to the functions $\left\{P_{\zeta}(V_{0}, \mu, p), n_s(\mu, p), \alpha_s(\mu, p), r(\mu, p)\right\}$.

%

We explored a number of methods (gradient boosted trees, two-layer fully connected neural network, Gaussian process regression) and found the k-nearest neighbour (KNN) method to work the best\footnote{Gaussian process regression with GPy \cite{gpy2014} also worked extremely well for the tree-level model but became more challenging to use with the full model.}. We used Scikit Learn's \cite{scikit-learn} built-in KNN function. We found the default settings of a leaf size of 30, Minkowski distance metric, and 10 distance-weighted neighbours to be close to optimal for all variants of both the tree-level and full model that we considered. For the tree-level model, we found essentially perfect agreement when testing the KNN model on holdout data.

A key step in order to get accurate results from the KNN method is to standardise the data. For the tree-level model, all parameters have support on a finite interval and hence simply performing a coordinate transformation such that all variables lie on a unit interval works well. We define
\begin{align}
z_{V_0} &=\frac{V_{0} -  \frac{1}{2}\left ( \rm{max}(V_{0}) + \rm{min}(V_{0})\right )}{ \rm{max}(V_{0}) - \rm{min}(V_{0})}  \nonumber \\
z_{\mu} &=\frac{\mu -  \frac{1}{2}\left ( {\rm max}(\mu) + \rm{min}(\mu)\right )}{ \rm{max}(\mu) - \rm{min}(\mu)} \nonumber \\ 
z_{p} &=\frac{p -  \frac{1}{2}\left ( {\rm max}(p) + {\rm min}(p)\right )}{ {\rm max}(p) - {\rm min}(p)} \, .
\end{align}
Since the relative entropy (and hence also the mutual information) is invariant under a bijective change of variables, this step also marginally simplifies our analysis of prior sensitivity, since now all variables have the same range.
Our strategy is to perform a KNN regression on the predictions obtained using the fiducial prior for model parameters. With the KNN algorithm in place there is no further need to solve the equations of motion in order to compute observables. The procedure is simply
\begin{itemize}
\item Take a random draw of the model parameters (for the tree level model this is simply $\{V_{0}, \mu, p\}$)
\item Substitute them into the KNN function $\left\{P_{\zeta}(V_{0}, \mu, p), n_s(\mu, p), \alpha_s(\mu, p), r(\mu, p)\right\}$
\item Repeat this procedure as many times as you like, since it is really fast
\end{itemize}
and hence it becomes trivial to study the impact for predictions of changing the prior on model parameters. 

\subsection{Large hierarchy in information loss between $\mu$ and $p$}
We start by computing the mutual information between each parameter and each observable for the tree-level model with fiducial priors. To do so we use the publicly available NPEET code~\cite{npeet, ver2013information}. Table \ref{table:MI_tree} shows that there are strong hierarchies in the mutual information. For instance, we see that $P_{\zeta}$ contains a great deal of information about $V_{0}$ while the other observables contain none at all. This is consistent with the previous discussion that for the tree-level model, $V_0$ drops out of the expressions for $\left\{n_s, \alpha_s, r \right\}$ provided the slow-roll expressions are valid.
\begin{table}[htbp]
\caption{Mutual information between each parameter and each observables for the fiducial tree-level model}
\begin{tabular}{@{}p{0.12\textwidth}*{4}{L{\dimexpr0.22\textwidth-2\tabcolsep\relax}}@{}}
\toprule
    & {$P_{\zeta}$} & {$n_s$} & {$\alpha_s$} & {$r$}  \\ 
 \midrule
    {$V_{0}$}  & 0.7 & 0.0 & 0.0 & 0.0  \\
    {$\mu$} & 0.1  & 0.2 & 0.1  & 0.1  \\ 
    {$p$} & 0.5  & 5.2 & 2.7  & 6.5  \\ 
\bottomrule
\end{tabular}
\label{table:MI_tree}
\end{table}

Somewhat more interesting is that table \ref{table:MI_tree} shows a strong hierarchy in information about $\mu$ and $p$. To explore this in more detail we study the effect of changing the prior. As discussed in \S\ref{sec:PGM_tree}, we do so by generating large samples of $\mu$ and $p$ drawn from Beta distributions with varying hyperparameters. We then feed these samples to the KNN learnt functions $\left\{ n_s(\mu, p), \alpha_s(\mu, p), r(\mu, p)\right\}$ to generate the corresponding distributions for observables. 

\begin{figure}[t]
  \centering
     \includegraphics[width=0.93\textwidth]{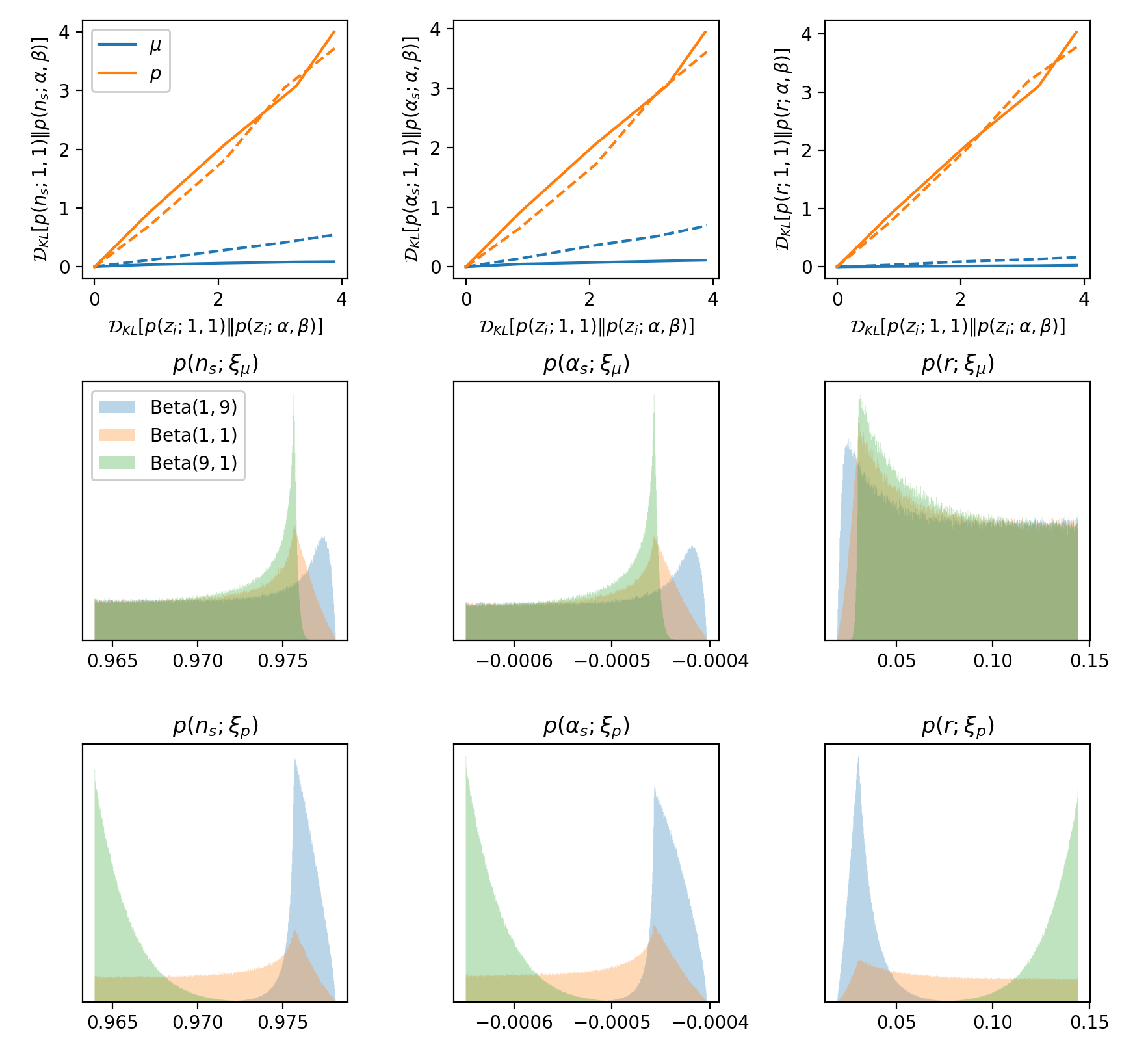}
\caption{Sensitivity of observables to changes in priors for $\mu$ and $p$: The top row shows the KL-divergence (relative entropy) of the distribution of observables $n_s$, $\alpha_s$ and $r$ against the KL-divergence of the prior of $z_\mu$ and $z_p$, as the hyperparameters move the priors away from the Beta(1,1) fiducial distribution. Solid lines represent changes in $\beta$ for fixed $\alpha=1$, and dashed lines changes in $\alpha$ for fixed $\beta=1$. The middle and bottom rows show the distributions for observables upon changes in the prior of $z_\mu$ (middle) and $z_p$ (bottom), particularly when the priors are described by the Beta distribution with $(\alpha=1, \beta=9)$, $(\alpha=1, \beta=1)$ and $(\alpha=9, \beta=1)$.}
  \label{fig:TL_prior_sens}
\end{figure}
As summarised in Fig.~\ref{fig:TL_prior_sens}, we find that the consequence of the hierarchy in information loss is that the model predictions are much more sensitive to changes of the prior on $z_p$ than to changes of the prior on $z_\mu$. 
The top row of plots compares the relative entropy of the distributions for observables to the relative entropy of the prior as a given hyperparameter is changed with respect to the fiducial distribution. We see that the relative entropy increases much more rapidly when the prior $z_p$ is changed, compared to when the prior on $z_\mu$ is changed. Indeed, referring back to the discussion surrounding Eq.~\eqref{eq:I_f_def}, the approximately linear slope with gradient close to one implies that the hierarchy is so strong that there is almost no information loss with respect to $p$ and substantial information loss with respect to $\mu$. The second and third row give example histograms for observables resulting from the extreme variations of the priors that we considered. Corroborating the takeaway message from the plots in the top row, we see that qualitatively the plots in the bottom row, which correspond to variations in the prior on $z_p$, appear to differ from one another significantly more than those in the middle row (which correspond to variations in the prior on $z_\mu$).
%
%

We conclude that to further refine the predictions of this model, one should focus model building efforts on understanding the microphysical considerations underlying $p$. Refining our understanding of $\mu$ however, will have little-to-no impact on the predictions of the model, since, with respect to $\mu$, the predictions are bordering on universal. 


This observation is rather timely, given the recent surge of interest in criteria like the WGC or the Swampland Distance Conjecture (SDC)~\cite{Ooguri:2006in}. These conjectures try to establish limits $\mu\lesssim \MC{O}(\Mp)$ as a consequence of consistency of a given EFT with general properties of candidate theories of quantum gravity like string theory. While the motivations for exploring these conjectures are diverse, one might have hoped that one result to come from this program would be more refined models of inflation (assuming inflation is allowed by whatever concrete statements might emerge from this program). Clearly, our results indicate that for models of axion monodromy inflation, this program is unlikely to be fruitful in this particular regard, in so much as the inflationary observables largely decouple from the precise value of $\mu$. Rather, we find that a potentially dramatic improvement in our ability to make predictions would follow from a better understanding of the physics underlying $p$. This highlights the potential value of the type of analysis we are proposing here. That the pursuit of more precise statements about predictions may be most effectively carried out through an iterative procedure of statistical analysis along the lines of this work, followed by directed model refinement based on the conclusions.

\paragraph{A comment on the lower bound on r} Having said that the model predictions largely decouple from $\mu$, it is important to note an exception --- the lower bound on the tensor-to-scalar ratio. This is a particularly important point to note at this stage, as a fantastic experimental effort is being made by ground experiments, both currently and in the near future, to further constrain the upper limit on $r$. As can be seen in Fig.~\ref{fig:TL_prior_sens} the distribution of $\mu$ changes the lower bound on $r$. While this change might be small, it can be important for the possibility of ruling out the model with future data. Indeed, while for the fiducial distribution the lower bound is $r \gtrsim 0.02$, for the distribution peaking at larger $\mu$ values $r \gtrsim 0.03$. It turns out that for the smallest values of $p$, $\mu$ is a good predictor of the lower bound on $r$ and $p$ decouples from the estimation of this bound. 
Another point we would like to make is that a na\"ive estimative of the value of $r$ using the approximations $V(\phi) \approx \phi^p$ and $\epsilon \propto 1/N_e$ leads to a predicted lower bound of $r > 0.007$ for the total number of e-folds $N_e=55$, which is far from the 0.02 bound we see from Fig.~\ref{fig:TL_prior_sens}. Indeed, for the values that $p$ and $\mu$ can take in our model, both parameters are important to determine $r$.

\subsection{The remarkable robustness of the distribution of the warp factor and $V_0$} \label{subsec:robustnessV0}
In this subsection we return to the discussion of deriving a prior on $V_0$ from top-down considerations. The main point we would like to emphasise is that simple observations arising from requiring consistency of the underlying physics can have strong ramifications for the form of the distribution. 

\begin{figure}[t]
  \centering
     \includegraphics[width=1.\textwidth]{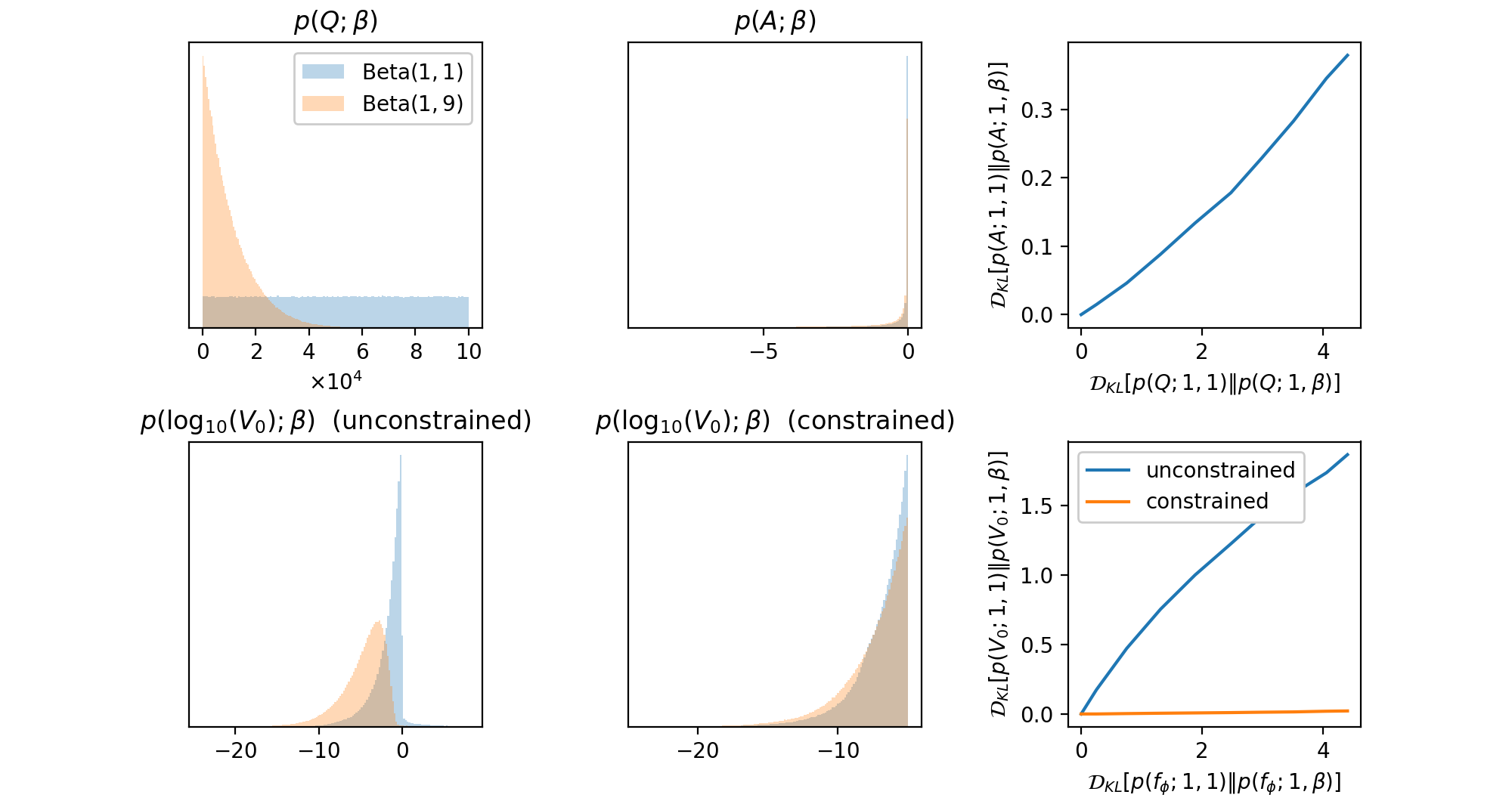}
\caption{Robustness of the distribution for $e^A$ and $V_0$: The top left plot shows the priors for $Q$ (and identically $M$) for different choices of the hyperparameter $\beta$. The top middle plot shows the resulting distribution for $A$ for both priors on $Q$ and $M$; we see how robust this is to changes on the prior. The top right plot shows the KL-divergence of the distribution of $A$ against the KL-divergence of the prior on $Q$ (and $M$), for changing $\beta$; we see there is high information loss. The bottom left plot shows the distribution of $\log_{10}(V_{0})$ before imposing the cut $V_0<10^{-5}$ for different priors on $f_\phi$ -- Beta(1,1) (blue) and Beta(1,9) (orange). The bottom middle plot shows the same distributions, but after imposing the cut $V_0<10^{-5}$. The bottom right plot shows the KL-divergence for the distribution of $V_0$ against the KL-divergence for the prior on $f_{\phi}$, for varying $\beta$. The blue line shows the unconstrained case, with little loss of information, and the orange line shows the case when the cut $V_0<10^{-5}$ has been imposed, with a large information loss.}
  \label{fig:warp_factor_and_V0_plots}
\end{figure}

\subsubsection{The distribution of the warp factor}
With our current knowledge, we cannot say much about the individual distributions of $Q$ and $M$ other than that they are integers in the range $[1,10^5]$. Given this large range we will simplify our analysis by approximating $Q$ and $M$ as continuous rather than discrete random variables. We will then, as before, use variation of the Beta distribution and computing the resulting relative entropy as our probe of prior sensitivity.

While we may not know much about the individual distributions of $Q$ and $M$, the constraints $QM < 10^5$ and $\exp(-Q/M) > 10^{-4}$ lead to dramatic information loss and hence a surprisingly robust prediction for the distribution of the warp factor. This point is illustrated in the top row of plots of Fig.~\ref{fig:warp_factor_and_V0_plots}, where the left plot shows two different possible distributions for $Q$ (and we use the same for $M$), and the middle plot shows the resulting distributions for $A = -Q/M$ (\emph{i.e.} the log of the warp factor) after enforcing the constraints. When we use the flat prior for $Q$ and $M$, the vast majority of samples get rejected by the bound $QM < 10^5$. For distributions more peaked towards lower values the rejection rate is lower, but if very peaked at low values the bound on $\exp(-Q/M) > 10^{-4}$ becomes increasingly important. These cuts on the distribution gives rise to significant information loss, and hence, by the data processing theorem, must result in the distribution of the warp factor being less sensitive to variations of the hyperparameter $\beta$ than $Q$ and $M$. This is confirmed by the third plot, which shows that the relative entropy of the warp factor is down by more than an order of magnitude compared to the relative entropy of $Q$ or $M$, for a given change in $\beta$.

In summary, applying strong constraints to otherwise broad distributions can be thought of as a particularly extreme form of data processing leading to information loss. Thus it can be possible to make robust statements about the form of the resulting distribution, despite significant uncertainty in the distribution of more fundamental quantities. 

This is reminiscent of Weinberg's arguments leading to an anthropic bound on the cosmological constant~\cite{Weinberg:1987dv}: The anthropic c.c. argument essentially relies on the fact that imposing the dramatically narrow anthropic cuts $-\rho_0\lesssim \Lambda\lesssim \MC{O}(10)\times \rho_0$ on an unknown prior probability distribution of $\Lambda$ with width $\MC{O}(\Mp)$ renders the 
distribution approximately flat in the range $-\rho_0\lesssim \Lambda\lesssim \MC{O}(10)\times \rho_0$. Moreover, this outcome is independent of the shape of the prior as long as this has no pole within the range of the anthropic cuts. Here $\rho_0\sim 10^{-120}\,\Mp^4$ denotes the magnitude of the present-day energy density of the universe. Clearly, Weinberg's observation exemplifies a very drastic example of data processing leading to information loss. 

The analogy with our situation is clear,  since we impose rather strong cuts on the range of allowed warp factor values as well as on the flux integer values by virtue of the tadpole constraint --- quantities which otherwise could take rather wide ranges. It is comforting that by numerically drawing $Q$ and $M$ we reproduce the analytical result for the distribution of the warp factor $e^A$ for the near-conifuld flux density computed in Refs~\cite{Denef:2004ze,Hebecker:2006bn}:
\be
{\rm p}(e^A)\sim \frac{1}{\ln\,e^A}\quad.
\ee


\subsubsection{The prior on $V_{0}$}
We can express $V_0$ as 
\be
V_0 = e^{4A(1-3/\ell)} f_{\phi}^{12/\ell} \, .
\ee
The exponential term is robust to changes in the distribution of $Q$ and $M$ as a direct consequence of the discussion of the previous subsection, so the question becomes, how does the prior on the axion decay constant $f_\phi$ affect the distribution of $V_{0}$? At first it appears that $V_0$ is sensitive to the distribution of $f_{\phi}$. Taking $f_\phi$ to be drawn from a rescaled Beta distribution on the range $[10^{-4}, 1]$, the bottom left plot of Fig.~\ref{fig:warp_factor_and_V0_plots} shows significant variation (baring in mind that it is the histogram of $\log_{10}(V_{0})$ shown). This point is perhaps more easily seen by the relative entropy plot on the bottom right (blue line), which shows a fairly modest degree of information loss. However, these plots neglect the constraint $V_0 < 10^{-5}$, which restricts us the tail of the distributions shown in the bottom left plot. After enforcing this constraint, we are left with the bottom centre plot, where the two histograms almost perfectly coincide. Indeed, our estimate of the relative entropy is consistent with these two distributions being identical (orange line in the bottom right plot).


%
%

\section{Analysis of the full model} \label{sec:results_full}

\begin{figure}[t]
  \centering
     \includegraphics[width=0.6\textwidth]{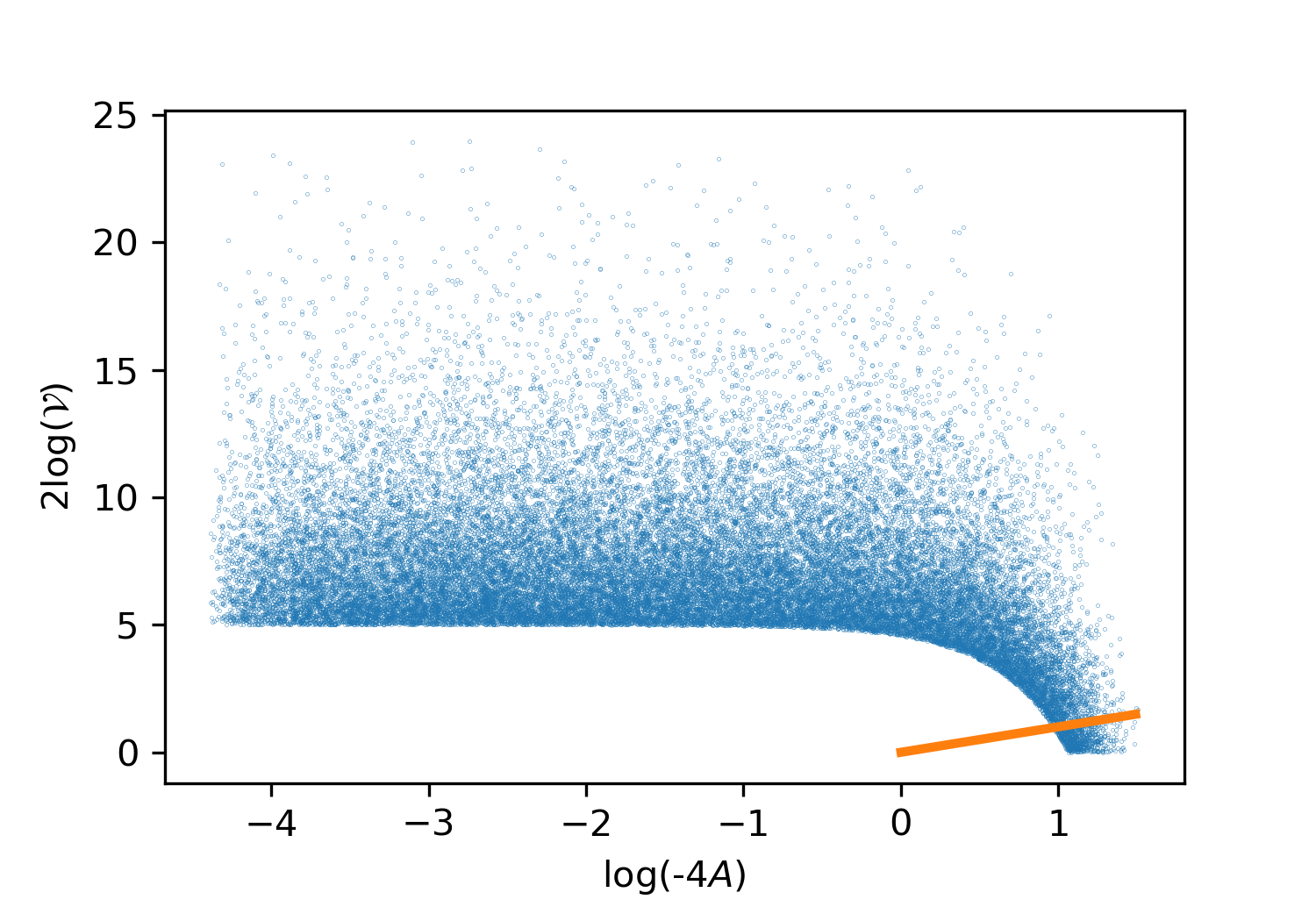}
  \caption{Scatter plot of $\log{(-4A)}$ versus $2\log{(\vol)}$. The orange line represents the region where $V_{\rm tree}$ is of the same order of magnitude as $V_{np}$ for $k=2$. The volume spans over a much wider range of magnitudes than the $\log$ of the warp factor, $A$. We can see that most realisations sit above the orange line and therefore lead to a negligible contribution from $V_{np}$} 
  \label{fig:non_perturbative_contrib}
\end{figure}

\subsection{Non-perturbative corrections}

\begin{figure}[t]
  \centering
     \includegraphics[width=0.6\textwidth]{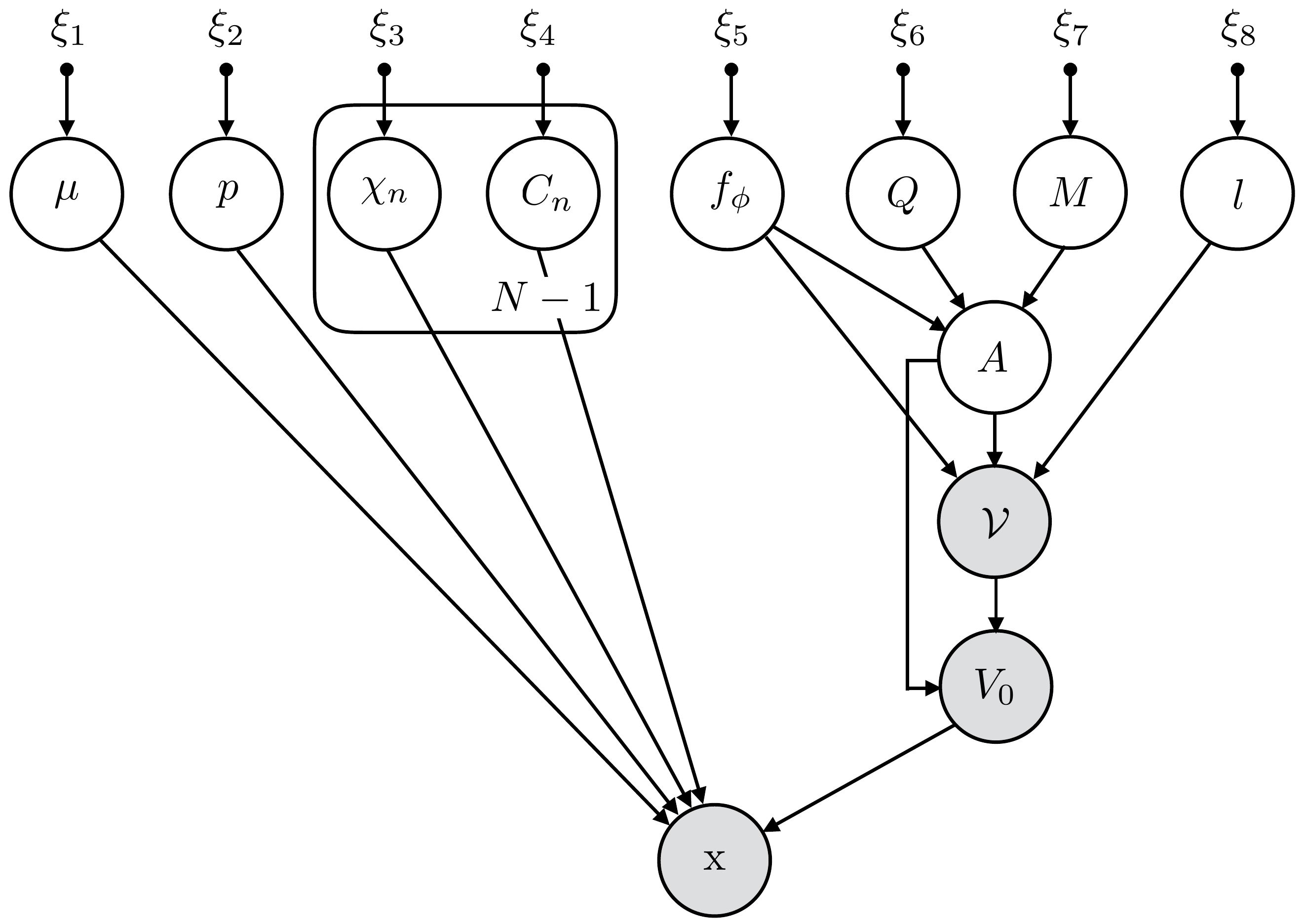}
  \caption{Graphical representation of $V_{\rm tree} + V_p$ as defined in Eq.~\eqref{eq:full_potential}}
  \label{fig:graph_Vp}
\end{figure}

According to Eq.~\eqref{eq:Corrections1} the non-perturbative corrections are exponentially suppressed by powers of the volume, which can take quite large values, so we can expect that for the majority of parameter space the effect of these contributions is negligible. On the other hand, if these corrections do become important, they can easily ruin otherwise successful inflation by trapping the inflation in a local minimum generated by the cosine potential. Thus for these corrections to allow monodromy inflation and yet be large enough to cause a cosine modulation of the tree level potential, a very delicate balance between parameters needs to occur. As a rough order of magnitude estimation of these relations, we can assume that the requirement for the corrections to be relevant is 
\begin{equation}
V_0 \sim \Lambda_{\rm UV}^4 e^{-S} \quad \rightarrow \quad 4A \sim \vol^k \quad .
\end{equation} 
The volume can take values spanning a range of magnitudes, making this relation hard to achieve: this is shown in Fig.~\ref{fig:non_perturbative_contrib}, where we look at the case of $k=2$. The vast majority of realisations sit above the orange $y=x$ line and lead to negligible contributions. Points below this line will easily spoil the inflaton potential. Given how unlikely it is to realise a potential modulated by the cosine corrections, we choose to focus our analises on the perturbative corrections only\footnote{It would be interesting to constrain the parameter space to the regions leading to interesting non-perturbative corrections and study phenomena such as features in the power spectrum and particle production, even though they are not generic predictions of our model. We leave this for future work.}. The potential we will now discuss is shown in Fig.~\ref{fig:graph_Vp} as a probabilistic graph.

\subsection{Perturbative corrections}

When we include perturbative corrections to the potential, the probabilistic nature of the model differs quite dramatically from the tree level case. The parameter $V_{0}$ no longer decouples from the predictions for $\{n_s, \alpha_s, r\}$ as in the tree-level case, on the contrary, it plays a crucial role in controlling the size of the perturbative correction. There are also two more parameters appearing in the action: $C_n$ and $\chi_n$. For simplicity we will study the case where $N=2$, and therefore $n$ only takes one value, $n=1$. While the probabilistic nature of the model changes, the relations between $f_\phi$, $Q$, $M$, $l$, $A$ and $\vol$ leading to the distribution of $V_0$ remain the same as discussed in \S\ref{subsec:robustnessV0}. For this reason, we will treat $V_0$, $\mu$, $p$, $\chi_n$ and $C_n$ as the model parameters. 

\begin{figure}[t]
  \centering
     \includegraphics[width=1.0\textwidth]{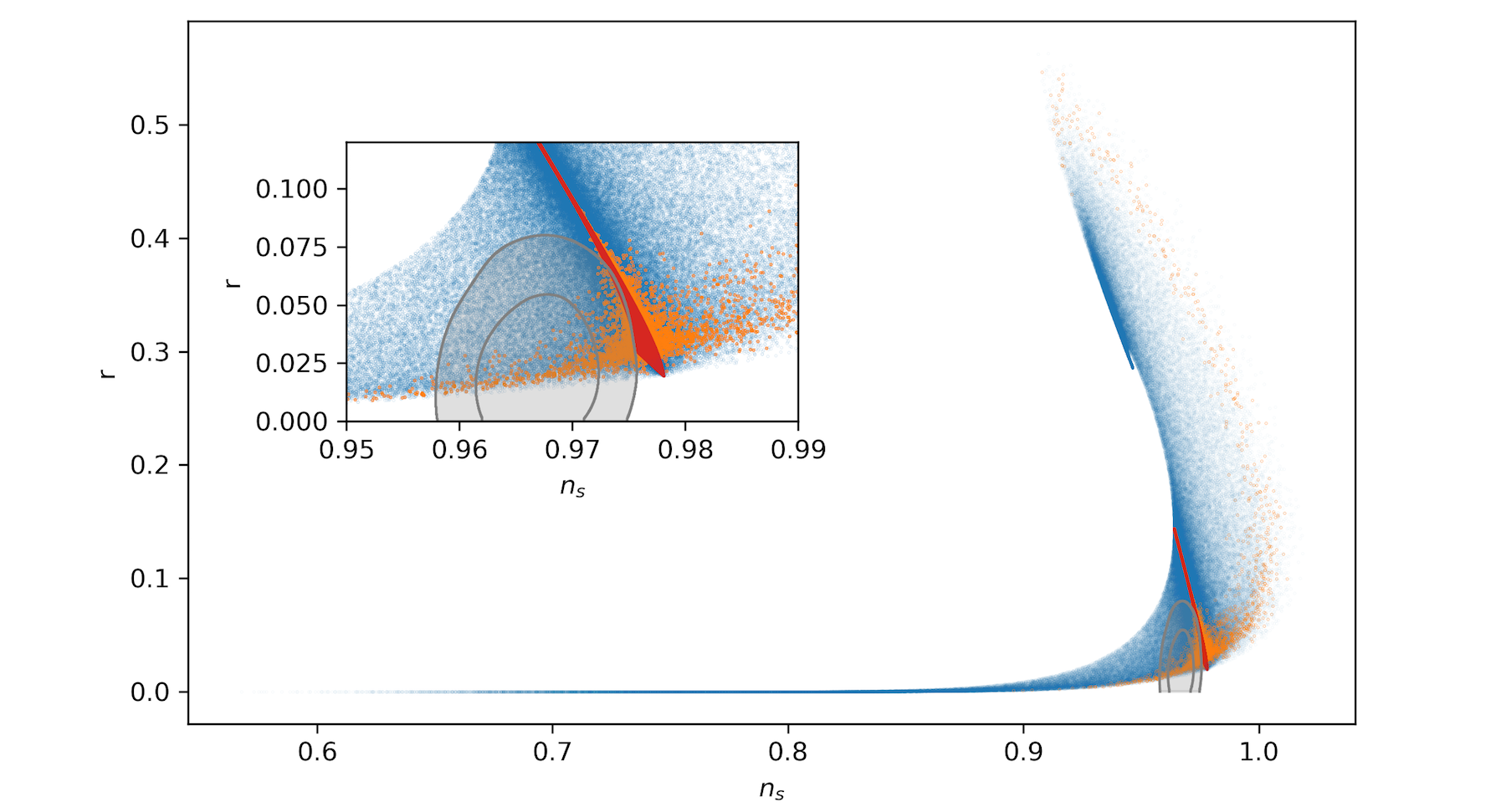}
\caption{Scatter plot of $n_s$ versus $r$: the red dots correspond to the predictions of the tree level model, the blue dots correspond to the predictions of the model when we include perturbative corrections, and the orange dots correspond to the subset of the latter which satisfy COBE normalisation. The grey contours are the $68\%$ and $95\%$ confidence limits from the 2018 Planck data~\cite{Akrami:2018odb}. We can see that the perturbative corrections have a very strong impact on the predictions from monodromy.}
  \label{fig:ns_r_comparison_of_tree_vs_full}
\end{figure}

Figure \ref{fig:ns_r_comparison_of_tree_vs_full} shows a scatter plot in the $n_s - r$ plane of samples generated by the fiducial tree-level model (red) and for the fiducial full model (blue), except with a flat prior on $\log_{10}V_{0}$ to better demonstrate the full range of the model. The orange points are those which satisfy constraints on the amplitude of the power spectrum, which we discuss in \S\ref{sec:conditioned_samples}. Clearly the range of possible values of $n_s$ and $r$ is much greater than the range of the tree level model and interestingly this includes a region which fits the Planck constraints better than the tree-level model. The multiple regions showing a denser clustering of blue points hints at the rich structure of the full model, which we will now explore in detail.

As with the tree-level case, we start by computing the mutual information between each parameter and each observable for the fiducial model. Table~\ref{table:MI_full} shows that in contrast to the tree-level model, now all observables contain some information about $V_{0}$. However $\{n_s, \alpha, r\}$ still contain significantly more information about $p$ than any other parameter. Strikingly, no observable seems to contain much information about the parameters $\chi_n$ and $C_n$. In other words, the predictions are universal with respect to the distribution of $\chi_n$ and $C_n$, despite the perturbative correction clearly having a large impact on the predictions.

\begin{table}[htbp]
\caption{Mutual information between each parameter and each observables for the fiducial full model}
\begin{tabular}{@{}p{0.12\textwidth}*{4}{L{\dimexpr0.22\textwidth-2\tabcolsep\relax}}@{}}
\toprule
    & {$P_{\zeta}$} & {$n_s$} & {$\alpha_s$} & {$r$}  \\ 
 \midrule
    {$V_{0}$}  & 0.7 & 0.3 & 0.2 & 0.2  \\
    {$\mu$} & 0.2  & 0.2 & 0.1  & 0.1  \\ 
    {$p$} & 0.1  & 0.8 & 0.9  & 0.6  \\ 
    {$\chi_n$} & 0.0  & 0.1 & 0.1  & 0.1  \\ 
    {$C_n$} & 0.0  & 0.1 & 0.1  & 0.1  \\ 
\bottomrule
\end{tabular}
\label{table:MI_full}
\end{table}

\begin{figure}[t]
  \centering
     \includegraphics[width=1.\textwidth]{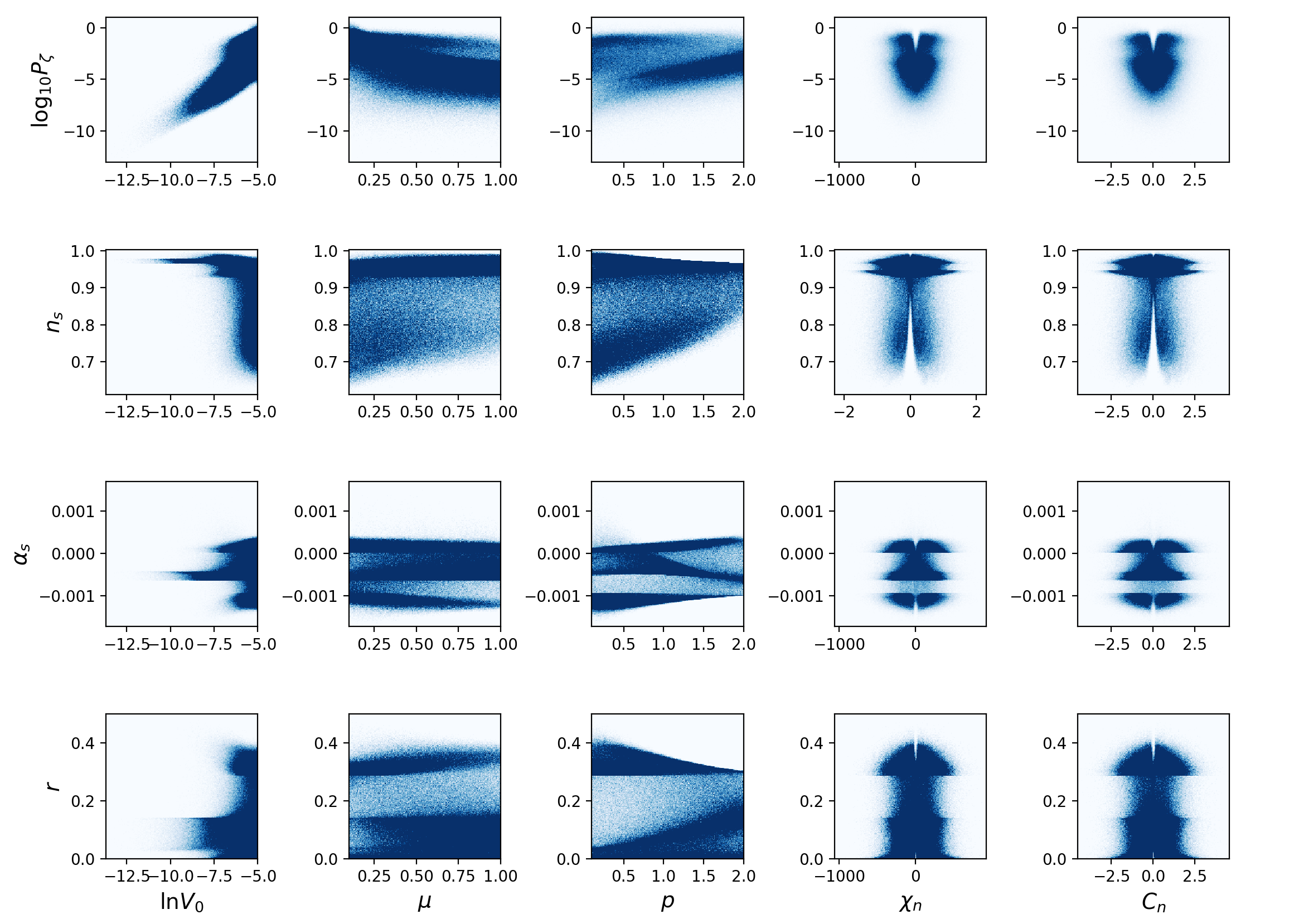}
\caption{2D histograms of observables versus model parameters}
  \label{fig:2d_hists_fiducial_full}
\end{figure}

\subsection{Sharp Transitions} \label{sec:AM_ST}
\begin{figure}[t]
  \centering
     \includegraphics[width=1.\textwidth]{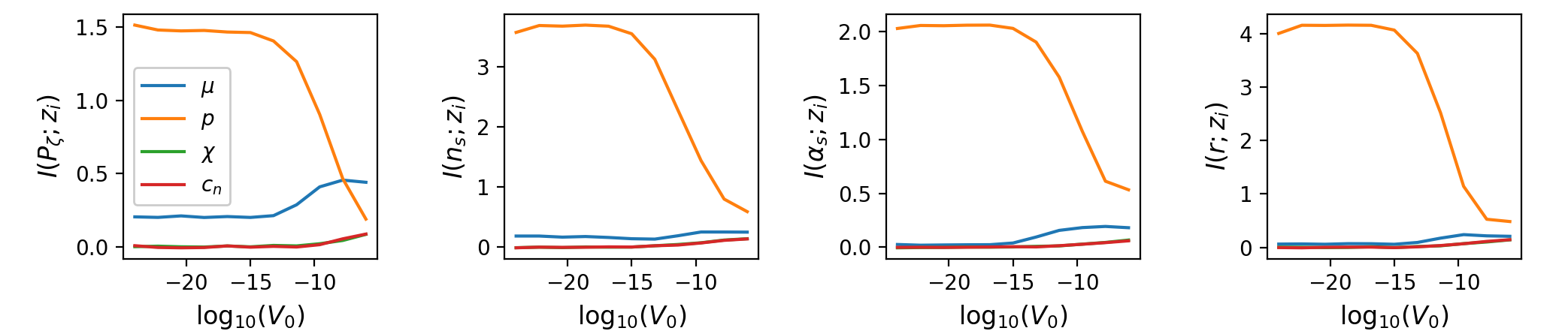}
\caption{Mutual information between observables and model parameters when conditioning on $\log(V_0)$. There is a sharp transition as $V_0$ gets large, which is particularly prominent for $p$. The sudden drop for $\log(V_0) \gtrsim -15$, implies $p$ stops being a good predictor.}
  \label{fig:MI_plots}
\end{figure}

We can glean a little more intuition from the fiducial model by making two--dimensional histograms for each observable and each parameter. 
Fig.~\ref{fig:2d_hists_fiducial_full} shows a number of sharp transitions in behaviour. We can think of a vertical slice of these plots as a (unnormalised) histogram of a given observable, conditioned on a particular value of the model parameter shown on the $x$-axis. For instance, we can think of a vertical slice of the top left plot as being proportional to the conditional probability density function p$(\ln P_{\zeta}\vert\ln V_{0})$. The first column therefore shows that as we vary $V_{0}$, there is a sharp transition, where for low values of $V_{0}$ all observables are much more predictive compared to when $V_{0}$ is large\footnote{Note that the normalisation is done over the 2D histograms. This means we don't necessarily see a dark blue, denser peak for regions which are highly predictive in the 1D sense of conditioning on the $x$-axis.}.
 This sharp transition in behaviour from highly predictive to far less predictive, is reminiscent of behaviour we saw when changing the mean of the random matrix example in \S\ref{subsec:sharp} (and is behaviour often seen in the context of high dimensional probability more generally), although here the underlying mechanism is rather different. In this case, the sharp transition occurs when the perturbative correction in Eq.~\eqref{eq:Corrections1} stops being negligible. When $V_{0}$ is small, this term is highly suppressed, and the model effectively reduces to the tree-level model. However, once $V_{0}$ reaches a critical value, this term quickly transitions to having a strong effect on the predictions of the model. Fig.~\ref{fig:MI_plots} shows the corresponding mutual information for these conditional probability plots. We see that when observables are conditioned on $V_{0}$ being small, the mutual information between the observable and the model parameter $p$ is very high, consistent with the discussion in \S\ref{sec:results_tree}, ultimately implying that in this regime $p$ is a good predictor for $n_s$. However, above a certain threshold value, the mutual information drops. In this regime, the hierarchies in parameter sensitivity are much less strong and the model is much less predictive. 

\subsection{Conditioning on satisfying observational constraints}\label{sec:conditioned_samples}

Remarkably, inflationary realisations consistent with observational constraints on the amplitude of the power spectrum live right on the edge of the sharp transition with respect to $V_0$ just described. This can be seen directly from the top left plot of Fig.~\ref{fig:2d_hists_fiducial_full}. Going back to Fig.~\ref{fig:ns_r_comparison_of_tree_vs_full}, the orange points are the subset of the full model sample which satisfy observational constraints on $P_{\zeta}$. It is clear that the perturbative correction continues to play a crucial role after conditioning on satisfying this constraint and that there is a cluster of points which also satisfy the observational constraints on $n_s$ and $r$. 

\begin{figure}[t]
  \centering
     \includegraphics[width=1.\textwidth]{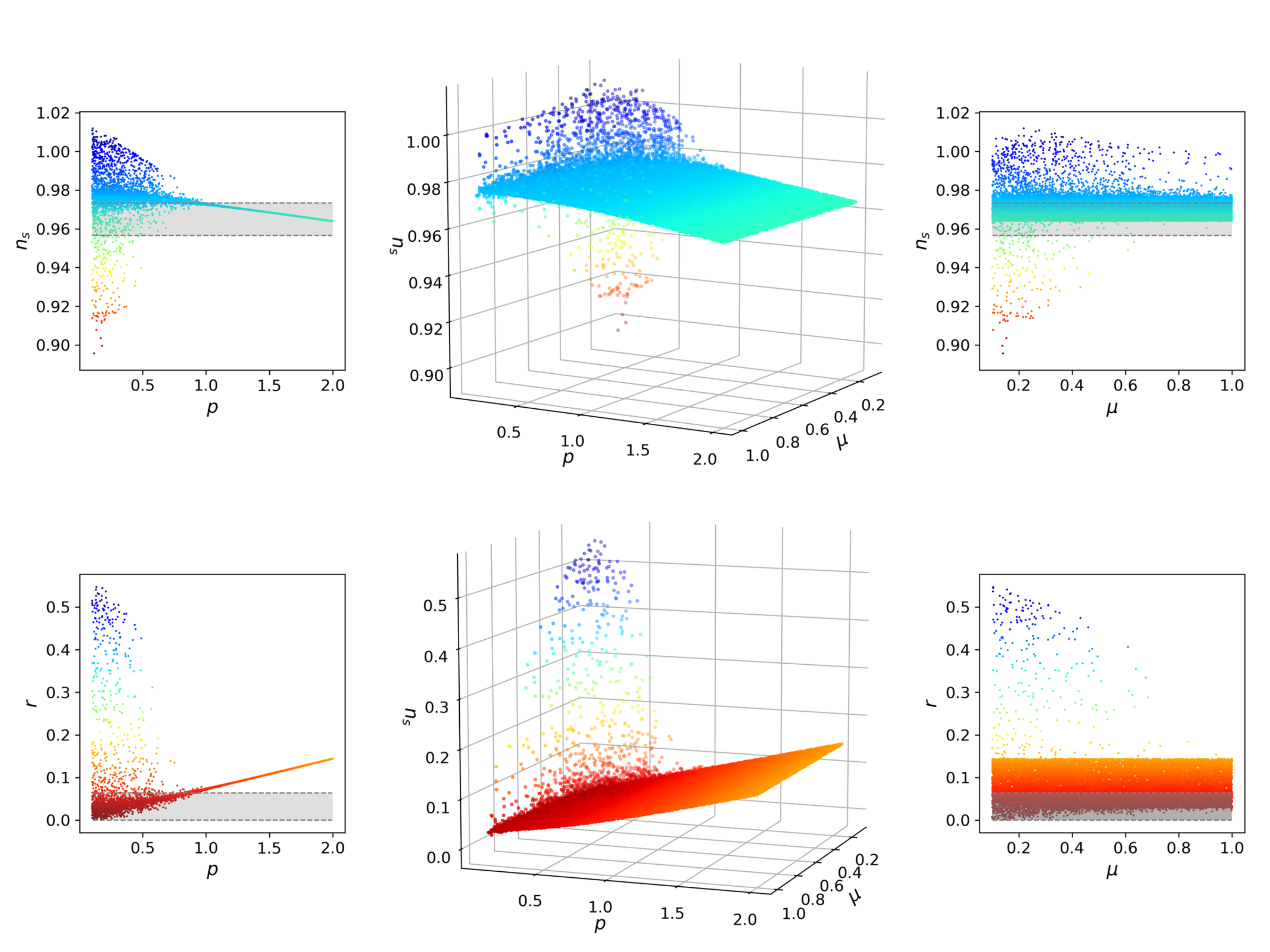}
\caption{Scatter plots of realisations conditioned to COBE normalisation: The central plots show $n_s$ (top) and $r$ (bottom) versus $p$ and $\mu$, and the side plots show the projections $n_s(p)$ (top left), $n_s(\mu)$ (top right), $r(p)$ (bottom left) and $r(\mu)$ (bottom right). The grey shaded areas correspond to the 95\% confidence limits from the Planck data~\cite{Akrami:2018odb}. The colour grading matches the $n_s$ (top) and $r$ (bottom) values. It is evident from the right hand side and central plots that perturbative corrections are important for small values of $p$, which are also imposed by constraints on $r$. So observational constraints force the model to be in the regime where perturbative corrections are important.}
  \label{fig:grid_ns_r_mu_p_scatter}
\end{figure}

To explore this in more detail, Fig.~\ref{fig:grid_ns_r_mu_p_scatter} shows scatter plots solely consisting of points that are consistent with the observational bounds on $P_{\zeta}$. The central plots are scatter plots of $\{n_s, \mu, p\}$ (top) and $\{r, \mu, p\}$ (bottom). The plots either side show the projections along the $\mu$ and $p$ axis and the grey regions between dashed lines show the 95\% confidence limits from the Planck results~\cite{Akrami:2018odb}. We can see that the transitional region discussed in \S\ref{sec:AM_ST} exhibits a particularly rich interplay between the parameters of the model and the observables $n_s$ and $r$. For large values of $p$, the points collapse onto a surface in the 3D plots. This effectively two-dimensional region corresponds to the tree-level model. However, for smaller values of $p$ we see points that do not sit on this surface, and are thus points determined also by the perturbative correction. We see that in order to satisfy constraints on $r$, the model must take small values of $p$. Thus, the key message to take away from these plots, and indeed, one of the main results of this paper with regard to the study of axion monodromy, is that compatibility with observational constraints coincides with the region of parameter space where perturbative corrections play a crucial role.

Indeed the effect of these corrections is almost self-contradictory: while they are not strong enough to destroy slow-roll inflation itself, they are easily capable of leading to drastic changes in observable predictions -- a state of affairs where these corrections can be described as being `dangerously irrelevant'~\cite{Silverstein:2017zfk} (see also Ref.~\cite{Silverstein:2008sg} where corrections of this size where labeled `neglibigle', \textit{i.e.} negligibly big). In addition, compatibility with observations forces the model to precisely live in the parameter space where the corrections can neither be considered completely irrelevant, nor are they completely dangerous.


\subsection{Prior sensitivity}

Finally, as with the tree level model, we test the sensitivity of predictions to variations of the prior for each model parameter. With this aim, just like in the tree-level case, we first make use of the KNN method to learn the mapping from parameters to observables. Again, we tried several methods and found this to be the most reliable. Once this mapping is determined, computing prior sensitivity is straightforward.

\begin{figure}[t]
  \centering
     \includegraphics[width=1.\textwidth]{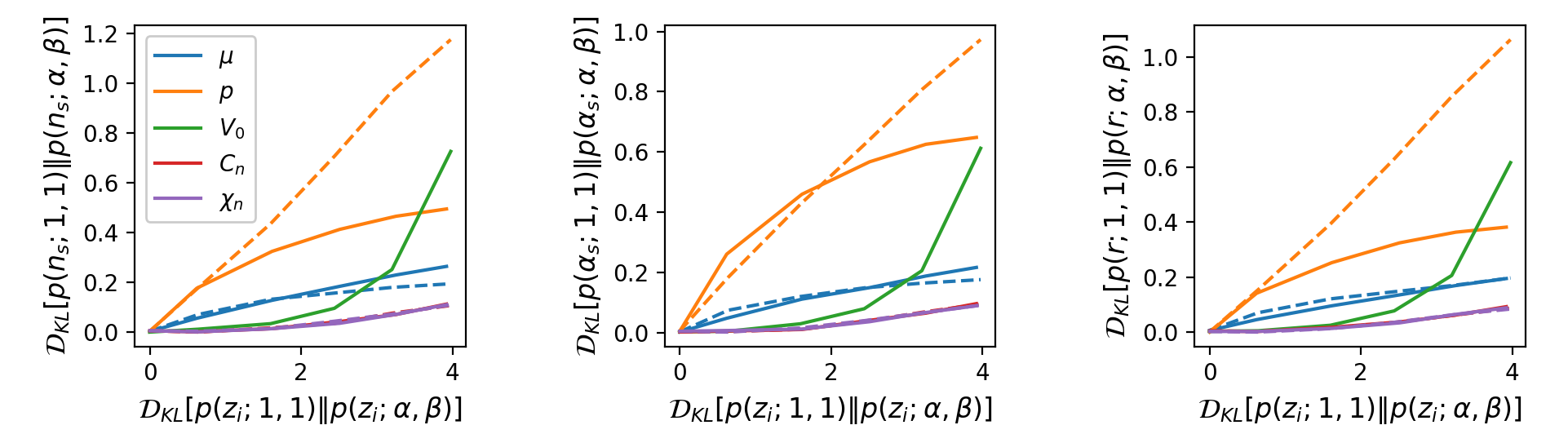}
\caption{Sensitivity of observables to changes in priors for model parameters: The plots show the KL-divergence (relative entropy) of the distribution of observables $n_s$ (left), $\alpha_s$ (middle) and $r$ (right) against the KL-divergence of the prior of $z_\mu$, $z_p$ $z_{V_0}$, $z_{C_n}$ and $z_{\chi_n}$, as the hyperparameters move the priors away from the Beta(1,1) fiducial distribution. Solid lines represent changes in $\beta$ for fixed $\alpha=1$ (towards smaller values), and dashed lines changes in $\alpha$ for fixed $\beta=1$ (towards larger values). For small deviations from the fiducial prior, the observables are particularly sensitive to $p$. This changes with the onset of the effect from perturbative corrections: there is a sharp rise in the sensitivity to $V_0$ as its distribution moves towards larger values, and there is a decrease of the sensitivity to $p$ when its distribution moves towards smaller values.}
  \label{fig:prior_sensitivity_full}
\end{figure}

Fig.~\ref{fig:prior_sensitivity_full} shows the relative entropy when changing the prior on each model parameter in turn. For small variations, it is clear that the prior on $p$ has the greatest effect. Dashed lines correspond to distributions peaking at high values, while solid lines represent distributions peaking at low values. The orange dashed line being steeper than the solid line is due to the effect of the perturbative correction being suppressed at large values of $p$ as discussed in the previous section. In contrast, the solid line becomes less steep due to the perturbative correction becoming more prominent. It is quite interesting to compare Fig.~\ref{fig:prior_sensitivity_full} with Fig.~\ref{fig:MI_plots}. While each figure teaches us something different about the model, in both cases we can interpret the plots as saying that we loose information about $p$ as the perturbative correction becomes larger (although, in Fig.~\ref{fig:MI_plots} the suppression was controlled by $V_{0}$, not by $p$). In, the case of Fig.~\ref{fig:MI_plots}, this is clearly achieved by a direct calculation of the mutual information, whereas Fig.~\ref{fig:prior_sensitivity_full} requires the use of the data processing theorem in order to appreciate the significance of the gradient of each line. 

Furthermore both figures~\ref{fig:MI_plots} and~\ref{fig:prior_sensitivity_full} allow us to detect the presence of a sharp transition. In the case of Fig.~\ref{fig:prior_sensitivity_full}, we see this through the sharp increase in the gradient of the green line, indicating that a sufficiently large change in the prior of $V_{0}$ leads to a sudden change in the predictions. 

Finally, we note the robustness of the predictions with regard to really quite large variations of the prior on $C_n$ and $\chi_n$. Given that the fiducial distribution for $V_{0}$ ensures that the perturbative corrections are playing a key role, one might be surprised at how unimportant it is to have detailed knowledge of the distribution of these parameters. This is perhaps the most striking demonstration of universality in this work.

\section{Conclusions} \label{sec:conclusions}
As model building in string theory continues to become increasingly sophisticated, it seems that the resulting models are becoming increasingly complex. One consequence of this is a tendency for a larger number of parameters appearing in the effective action, typically leading to a broader range of possibilities for observables. In addition to this, it is clear that any one model constructed in string theory harbours significant theoretical uncertainty. Two questions then arise. 
\begin{enumerate}
\item Given the potentially very broad range of possible observable outcomes in a given model, how does one best make use of observational data in guiding the refinement of the model?
\item How can one reliably make predictions for future surveys?
\end{enumerate}
In this work we have proposed that by incorporating statistical methods into the model building process, we can make progress with both of these questions. While this point may seem obvious (and far from original), the reason we are excited about this approach now and the reason we believe this approach is promising, essentially boils down to two observations.
\begin{enumerate}
\item For models derived from string theory, parameters appearing in the effective action can be subject to constraints which place strong limitations on the prior of that parameter.
\item Inflation may be regarded as an information bottleneck between the UV theory and observations. This bottleneck will necessarily result in some notion of universality in the model predictions.
\end{enumerate}
By better understanding the probabilistic nature of the model and hence the conditions under which predictions are robust, we can hope to make progress with regard to both questions 1 and~2. 

With regard to question question 1, by knowing if a model exhibits hierarchies in parameter sensitivity, in trying to refine the model we can then go on to prioritise those parameters to which observables are most sensitive. In the case of axion monodromy inflation studied in this work, it is clear that $p$ and $V_{0}$ play especially important roles in understanding the predictions of the model. Thus, future work seeking to refine the predictions should focus on the physics underlying $p$, such as flattening mechanisms, or refine model building considerations relating to $V_{0}$ such as improving our understanding of perturbative corrections. In contrast, refining the modelling of, say, the parameter $\mu$ (\emph{e.g.} via obtaining a quantitative version of the weak gravity conjecture), does not seem likely to be so fruitful. 

Regarding question 2, by classifying the model building assumptions in this way, we can hope to find ourselves in a situation where certain ``known unknowns'' can be shown to be unimportant. For instance, this could have been the case with the perturbative corrections in axion monodromy, had the conclusion of our analysis been that they are generically extremely suppressed. In reality, we found the opposite. We found that consistency with current constraints implies working in a region of parameter space where such corrections are extremely important. It is striking that we seem to have found ourselves in a situation where such corrections are important for making predictions and yet do not dominate to the extent of spoiling the possibility of large-field inflation occurring. In this precise sense we seem to have found direct evidence for the `dangerous irrelevance' of string theory and its corrections~\cite{Silverstein:2008sg,Silverstein:2017zfk}.

That said, there is much cause for optimism. While the perturbative corrections clearly are important, they are also controllable. While we did not make a full analysis of next to leading order terms and higher, there is good reason to think that only a finite number of terms will be important for understanding the predictions of the model.


\subsection*{Future Directions}

In addition to further developing our understanding of axion monodromy, almost all aspects of the method used in this work could be developed much further. This project draws on ideas from information geometry, machine learning, Bayesian statistics and high dimensional probability and we barely scratched the surface of any of these fields. As such, this work could be extended in quite a few directions.

For example, the inflation model we studied was in no sense high-dimensional but we hope the methods proposed here will become invaluable when studying models involving either a large number of fields or a large number of parameters. Some of the challenges and opportunities one would likely encounter when exploring such a model include the following.
\begin{itemize}
\item In this work we used rudimentary machine learning methods to learn the \emph{deterministic} mapping between model parameters and observables. We considered a number of approaches but ultimately settled for one of the simplest, namely a k-nearest neighbours algorithm. It seems unlikely that this will be the best way of handling a model with a very large number of parameters. It is likely that other more sophisticated machine learning algorithms will perform better in a higher-dimensional setting. Another possibility is to rather than try to learn the deterministic mapping, to fully embrace the probabilistic nature of the model and instead try to learn the \emph{stochastic} mapping between hyperparameters and distributions of observables. This would essentially replace a high dimensional but deterministic problem with a lower dimensional probabilistic one.
\item We embraced certain notions from the field of high-dimensional probability, such as universality and sharp transitions, but we made no attempt to actually make use of the many results that exist on this topic. This is of course in part because our model was in no sense high-dimensional but more generally we decided to take a purely numerical approach, making no attempt at analytical studies of our probabilistic model. It seems likely that in a truly high dimensional model there would be an opportunity to gain analytic control, and robust results, using mathematical concepts that have not seen a great deal of application in string theory to date.
\end{itemize}

We made a start in exploring information loss using tools from information theory. Our primary tools were the mutual information and relative entropy. It is clear however, that there is far more that could be done in this direction. Information theory and information geometry are of course rich fields and many of the key results of recent years have not been taken advantage of in this work. For instance, we made no attempt to bound the relative entropy or any other $f$-divergence in our model, and yet a large literature exists on the subject.

Finally, a very natural next step would be to perform Bayesian inference. Universality is in a sense a mixed blessing in so much as the more robust the predictions, the less we can learn about the underlying parameters from observation. Thus, in the case of axion monodromy inflation studied here, it is clear that we can infer more about $V_{0}$ and $p$ than $\mu$, $C_n$ and $\chi_n$. Performing Bayesian inference in this context, combined with sensitivity analysis along the lines of what we have done here, would ultimately be the most rigorous way to address question 1.

\section*{Acknowledgements}

JF would like to thank Kepa Sousa and Layne Price for many extremely valuable discussions during the very early stages of this work. We also owe special thanks to Boris Leistedt for advice relating to the machine learning aspect of this project. In addition, we benefited significantly from discussions with Thomas Bachlechner, Stefano Di Vita, Richard Easther, Thomas Edwards, Nemanja Kaloper, M.C. David Marsh, Liam McAllister, Hiranya Peiris and John Stout.
JF and AW are supported by the ERC Consolidator Grant STRINGFLATION under the HORIZON 2020 grant agreement no. 647995. MD is in part supported by the German Science Foundation (DFG) within the Collaborative Research Centre 676 \textit{Particles, Strings and the Early Universe}, and in part by the ERC Consolidator Grant STRINGFLATION under the HORIZON 2020 grant agreement no. 647995.

\section*{Appendix}
\appendix
\section{Proof of the Data Processing Theorem}\label{sec:DPT_proof}

In this paper we are interested in applying the Data Processing Theorem to situations where the dimensionality of the model's parameter space is greater than the dimensionality of the observable space and the parameters are statistically independent (and potentially also identically distributed). For this reason, in this appendix we give a proof of the data processing theorem suitable for our needs. 

Consider a channel $\kappa$ that produces $x$ given the vector ${\bf z}:=(z_{1},\dots, z_{n})$. We wish to compare the $f$-divergence of two distributions $D_{f}(p({\bf z})\Vert q({\bf z}))$ to the $f$-divergence of the corresponding distributions output from the channel $D_{f}(p_{\kappa}(x)\Vert q_{\kappa}(x))$. First, note that the $f$-divergence of the joint distributions $p(x,{\bf z})$ contains no more information than the distribution $p({\bf z})$ and hence $D_{f}(p(x,{\bf z})\Vert q(x,{\bf z})) = D_{f}(p({\bf z})\Vert q({\bf z}))$. This can be seen explicitly by writing
\begin{align}
D_{f}(p(x,{\bf z})\Vert q(x,{\bf z})) 
&:=  \mathbb{E}_{q(x,{\bf z})} \left [ f\left (\frac{p(x,{\bf z})}{q(x,{\bf z})} \right )\right ] \nonumber \\ 
& = \mathbb{E}_{q(x,{\bf z})} \left [ f\left (\frac{p(x|{\bf z})p({\bf z})}{q(x|{\bf z})q({\bf z})} \right )\right ] \nonumber \\
& = \mathbb{E}_{q({\bf z})} \left [ f\left (\frac{p({\bf z})}{q({\bf z})} \right )\right ] \nonumber \\
& =  D_{f}(p({\bf z})\Vert q({\bf z})) \, .
\end{align}
Next, we perform chain rule again but this time pulling out a factor of $p(x)$ and then applying Jensen's inequality
\begin{align}
D_{f}(p(x,{\bf z})\Vert q(x,{\bf z})) 
&:=  \mathbb{E}_{q(x)}\mathbb{E}_{q({\bf z}|x)} \left [ f\left (\frac{p(x,{\bf z})}{q(x,{\bf z})} \right )\right ] \nonumber \\ 
& \geq \mathbb{E}_{q(x)} \left [ f\left (\mathbb{E}_{q({\bf z}|x)}\left[ \frac{p(x,{\bf z})}{q(x,{\bf z})}\right ] \right )\right ] \nonumber \\ 
& = \mathbb{E}_{q(x)} \left [f\left(\frac{p(x)}{q(x)}\right )\right ] \nonumber \\
& = D_{f}(p(x)\Vert q(x)) \, .
\end{align}
This completes the usual proof of the data processing inequality, however we are interested in the special case where the $z_{i}$ are independent. At least for the case of the relative entropy, the divergence is additive. That is, for 
\be
p({\bf z}) = \prod_{1}^{n}p(z_i)\, ,
\ee
we have
\begin{align}
D_{\rm KL}(p({\bf z})\Vert q({\bf z})) = \sum_{i}^{n} D_{\rm KL}(p(z_i)\Vert q(z_i))\, ,
\end{align}
and hence
\be
\sum_{i}^{n} D_{\rm KL}(p(z_i)\Vert q(z_i)) \geq D_{\rm KL}(p(x)\Vert q(x)) \, .
\ee

\section{Universality in RMT from a high dimensional probability perspective}\label{app:HDPRMT}

To understand universality from a high probability perspective, we start with an informal definition given in Ref.~\cite{van2014probability}, where universality is stated as
\begin{quote}
If $z_{1},\dots,z_{n}$ are independent (or weakly dependent) random variables, then the expectation $\mathbb{E}[f(z_{1},\dots,z_{n})]$ is "insensitive" to the distribution of $z_{1},\dots,z_{n}$ when f is "sufficiently smooth".
\end{quote}
The meaning of "insensitive" and "sufficiently smooth" can be made precise in a number of ways. Perhaps the simplest version follows from the Lindeberg method, and gives rise to the following theorem
\begin{quote}
Let $\bf y$ and $\bf z$ be random vectors in $\mathbb{R}^{n}$ with independent coordinates. Then for any $f: \mathbb{R}^{n}\rightarrow \mathbb{R}$, the distribution of $f$ shows universality if 
\begin{align}\label{eq:universality}
\left |\mathbb{E}[f({\bf y})] - \mathbb{E}[f({\bf z})]\right | &\leq \sum^{n}_{i=1}\left\|\frac{\partial f}{\partial y_{i}}\right\|_{\infty}\left |\mathbb{E}[y_{i}] - \mathbb{E}[z_{i}]\right | \nonumber\\
&+ \frac{1}{2}\sum^{n}_{i=1}\left\|\frac{\partial^{2} f}{\partial y_{i}^{2}}\right\|_{\infty}\left |\mathbb{E}[y_{i}^{2}] - \mathbb{E}[z_{i}^{2}]\right | \nonumber\\ 
&+ \frac{1}{6}\sum^{n}_{i=1}\left\|\frac{\partial^{3} f}{\partial y_{i}^{3}}\right\|_{\infty}\mathbb{E}[|y_{i}|^{3} - |z_{i}|^{3}] \, .
\end{align}
\end{quote}
Specifically, if $y_i$ and $z_i$ have the same mean and variance, the last term defines the smoothness of $f$ requirement through its third derivative. 
It should be noted that a number of more refined theorems exist, including versions which do not require independence of $z_{1},\dots,z_{n}$  \cite{van2014probability}.

A particularly striking demonstration of these basic principles of high dimensional probability and their usefulness arrises in the study of large random matrices. Let us look at the global statistics of Wigner matrices, as discussed in \S\ref{subsec:universality}. As discussed, the spectrum of eigenvalues of an $n \times n$ Wigner matrix, as $n$ becomes large and regardless of the choice of prior on the entries of the matrix, converges to the Wigner semicircle law:
\be
p(\lambda) = \frac{1}{2\pi}\sqrt{4-\lambda^2}\delta_{|\lambda|\leq 2}  .
\ee
This is illustrated by the bottom right hand plot of Fig.~\ref{fig:prior_sens_uni}. There are numerous ways in which this result has been proven, which we will not discuss here. Instead we will sketch a strategy used in high dimensional probability, which is commonly applied to many universality results. 
To obtain the form of the limiting distribution often one takes the following two step procedure:
\begin{enumerate}
\item Show that the distribution is universal in the large $n$ limit by, for instance, showing that Eq.~\eqref{eq:universality} applies.
\item Compute the distribution by assuming that the prior takes a particularly simple form which enables an explicit calculation. 
\end{enumerate}
The first step is not obvious in so much as it may not be immediately clear how the expectation values in Eq.~\eqref{eq:universality} can be used to show that an entire function takes a universal form. To see this, note that one can define the distribution of eigenvalues as the average histogram of eigenvalues with infinitesimal binning. In other words, $p(\lambda)$ is the average sum of a set of "spikes" at locations $\lambda$, which can be formally written in terms of the expectation value of the sum of point masses:
\be
\mu_n \equiv \mathbb{E}\left [\frac{1}{n}\sum_{i=1}^{n}\delta_{\lambda_{i}} \right ]\, .
\ee
It is possible to show that this quantity exhibits universality in the large $n$ limit using Eq.~\eqref{eq:universality} \cite{van2014probability}. 

For the second step, the most common choice of prior is the Gaussian distribution, which is in many cases sufficiently simple to permit analytical calculations even for finite $n$. Indeed, when the prior of the entries $(M_n)_{ij}$ is Gaussian, the entries of $A_{n}$ are statistically independent and at the same time the ensemble is invariant under unitary transformations. This special subclass of Wigner matrices known as the Gaussian Unitary Ensemble (GUE) allows a fantastic degree of analytic control. Let us look again at the histograms in Fig.~\ref{fig:prior_sens_uni}. While the analytic form for the density functions of the second row are not known, for the first row which represents the GUE at finite $n$, we simply have 
\be
p(x) = \frac{1}{n\sqrt{2\pi}}e^{-x^{2}/2}\sum^{n-1}_{j = 0}\frac{H_{j}^{2}(x/\sqrt{2})}{2^{j}j!} \ ,
\ee
where we define $x = \lambda/\sqrt{n}$ and $H_{j}(x)$ are Hermite polynomials (see, for instance, Ref.~\cite{livan2018introduction} for a detailed derivation). In the large $n$ limit this function converges to the Wigner semi-circle law.

\bibliographystyle{JHEP}

\bibliography{references}

\end{document}